\pdfoutput=1
\newif\ifDraft\Drafttrue

\documentclass[letterpaper,twocolumn,10pt]{article}
\usepackage{usenix}
\usepackage{url}
\usepackage{graphicx}
\usepackage{natbib}
\setcitestyle{numbers,square}

\begin{document}

\date{}
 
\title{The Ideal Versus the Real: Revisiting the History of Virtual Machines and Containers}

\author{Allison Randal, University of Cambridge}

\maketitle

\begin{abstract}
The common perception in both academic literature and the industry
today is that virtual machines offer better security, while containers
offer better performance. However, a detailed review of the history of
these technologies and the current threats they face reveals a
different story. This survey covers key developments in the evolution
of virtual machines and containers from the 1950s to today, with an
emphasis on countering modern misperceptions with accurate historical
details and providing a solid foundation for ongoing research into the
future of secure isolation for multitenant infrastructures, such as
cloud and container deployments.
\end{abstract}

\section{Introduction} \label{sec-intro}

Many modern computing workloads run in multitenant environments, such as cloud or containers, where each physical machine is split into hundreds or thousands of smaller units of computing, called virtual machines, containers, cloud instances, or more generically \textit{guests}. Typically, a single \textit{tenant} (a user or group of users) is granted access to deploy guests in an orchestrated fashion across a cloud or cluster made up of hundreds or thousands of physical machines located in the same data center or across multiple data centers, to facilitate operational flexibility in areas such as capacity planning, resiliency, and reliable performance under variable load. Each guest runs its own (often minimal) operating system and application workloads, and maintains the illusion of being a physical machine, both to the end users who interact with the services running in the guests, and to developers who are able to build those services using familiar abstractions, such as programming languages, libraries, and operating system features. The illusion, however, is not perfect, because ultimately the guests do share the hardware resources (CPU, memory, cache, devices) of the underlying physical host machine, and consequently also have greater access to the host's privileged software (kernel, operating system) than a physically distinct machine would have.

Ideally, multitenant environments would offer strong isolation of the guest from the host, and between guests on the same host, but reality falls short of the ideal. The approaches that various implementations have taken to isolating guests have different strengths and weaknesses. For example, containers share a kernel with the host, while virtual machines may run as a process in the host operating system or a module in the host kernel, so they expose different attack surfaces through different code paths in the host operating system. Fundamentally, however, all existing implementations of virtual machines and containers are leaky abstractions, exposing more of the underlying software and hardware than is necessary, useful, or desirable. New security research in 2018 delivered a further blow to the ideal of isolation in multitenant environments, demonstrating that certain hardware vulnerabilities related to speculative execution---including Spectre, Meltdown, Foreshadow, L1TF, and variants---can easily bypass the software isolation of guests.

Because multitenancy has proven to be useful and profitable for a large sector of the computing industry, it is likely that a significant percentage of computing workloads will continue to run in multitenant environments for the foreseeable future. This is not a matter of na{\"i}vet{\'e}, but of pragmatism: these days, the companies who provide and make use of multitenant environments are generally fully aware of the security risks, but they do so anyway because the benefits---such as flexibility, resiliency, reliability, performance, cost, or any of a dozen other factors---outweigh the risks for their particular use cases and business needs. That being the case, it is worthwhile to take a step back and examine how the past sixty years of evolution led to the current tension between secure ideals and flawed reality, and what lessons from the past might help us build more secure software and hardware for the next sixty years.

\begin{figure*}
  \includegraphics[width=\textwidth]{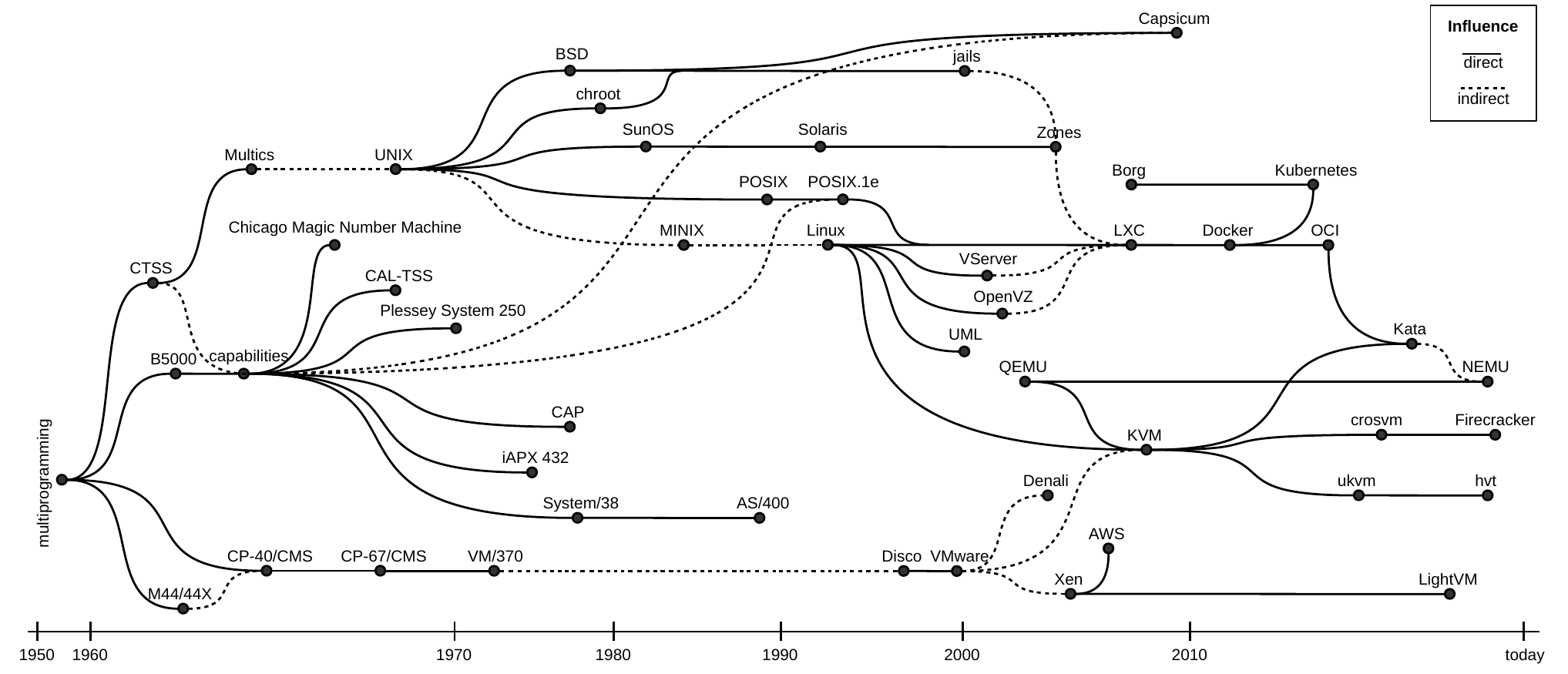}
  \caption{The evolution of virtual machines and containers.}
  \label{fig:timeline}
\end{figure*}

This survey is divided into sections following the evolutionary paths of the technologies behind virtual machines and containers, generally in chronological order, as illustrated in Figure \ref{fig:timeline}. Section \ref{sec-common-origins} explores the common origins of virtual machines and containers in the late 1950s and early 1960s, driven by the architectural shift toward multitasking and multiprocessing, and motivated by a desire to securely isolate processes, efficiently utilize shared resources, improve portability, and minimize complexity. Section \ref{sec-early-vms} examines the first virtual machines in the mid-1960s to 1970s, which primarily aimed to improve resource utilization in time-sharing systems. Section \ref{sec-capabilities} delves into the capability systems of the early 1960s to 1970s---the precursors of modern containers---which evolved along a parallel track to virtual machines, with similar motivations but different implementations. Section \ref{sec-modern-vms} outlines the resurgence of virtual machines in the late 1990s and 2000s. Section \ref{sec-containers} traces the emergence of containers in the 2000s and 2010s. Section \ref{sec-security} investigates the impact of recent security research on both virtual machines and containers. Section \ref{sec-related} briefly looks at the relationship between virtual machines and containers and the related terms ``cloud'', ``serverless'', and ``unikernels''.

\section{Terminology} \label{sec-terminology}

For the sake of clarity, this survey consistently uses certain modern or common terms, even when discussing literature that used various other terms for the same concepts.

\begin{itemize}
  \item \textbf{container}: The term ``container'' does not have a single origin, but some early relevant examples of use are Banga \textit{et al.} \cite{banga_resource_1999} in 1999, Lottiaux and Morin \cite{lottiaux_containers:_2001} in 2001, Morin \textit{et al.} \cite{morin_towards_2002} in 2002, and Price and Tucker \cite{price_solaris_2004} in 2004. Early literature on containers confusingly referred to them as a kind of virtualization \cite{price_solaris_2004, soltesz_container-based_2007, miller_exercise_2012, joy_performance_2015, catuogno_evaluation_2016, claassen_linux_2016}, or even called them virtual machines \cite{soltesz_container-based_2007}. As containers grew more popular, the confusion shifted to virtual machines being called containers \cite{bratus_vm-based_2010, zhai_cqstr:_2016}. This survey uses the term ``container'' for multitenant deployment techniques involving process isolation on a shared kernel (in contrast with \textit{virtual machine}, as defined below). However, in practice the distinction between containers and virtual machines is more of a spectrum than a binary divide. Techniques common to one can be effectively applied to the other, such as using system call filtering with containers, or using seccomp sandboxing or user namespaces with virtual machines.

  \item \textbf{complexity}: There are many dimensions to complexity in computing, but in the context of multitenant infrastructures some uniquely relevant dimensions are keeping each guest, the interactions between guests, and the host's management of the guests as small and simple as possible. The implementation technique of isolation supports minimizing complexity by restricting access to internal knowledge of the guests and host, and providing well-defined interfaces to reduce the complexity of interactions between them.

  \item \textbf{guest}: The term ``guest'' had some early usage in the 1980s for the operating system image running inside a virtual machine \cite{nanba_vm/4:_1985}, but was not common until the early 2000s \cite{waldspurger_memory_2002, barham_xen_2003}. This survey uses ``guest'' as a general term for operating system images hosted on multitenant infrastructures, but occasionally distinguishes between virtual machine guests and container guests.

  \item \textbf{kernel}: A variety of different terms appear in the early literature, including ``supervisory program'' \cite{codd_multiprogramming_1959}, ``supervisor program'' \cite{amdahl_architecture_1964}, ``control program'' \cite{nelson_mapping_1964, opler_multiprogramming:_1959, adair_virtual_1966}, ``coordinating program'' \cite{opler_multiprogramming:_1959}, ``nucleus'' \cite{buzen_evolution_1973, noauthor_control_1971}, ``monitor'' \cite{wilkes_time-sharing_1968}, and ultimately ``kernel'' around the mid-1970s \cite{lipner_security_1974, popek_verifiable_1975}. This survey uses the modern term ``kernel''.

  \item \textbf{performance}: There are many dimensions to performance in computing, but in the context of multitenant infrastructures some uniquely relevant dimensions are the performance impact of added layers of abstraction separating the guest application workload from the host, balanced against the performance benefits of sharing resources between guests and reducing wasted resources from unused capacity. At the level of a single machine, this involves running multiple guests on the same machine at the same time, with potential for intelligent, dynamic scheduling to extract more work from the same resource pool. Across multiple machines this involves a larger pool of shared resources, more flexibility to balance work, and options for hetrogenous hardware with resource-affinity configurations (e.g. a mixture of some CPU-heavy machines and some storage-heavy machines, with workload allocation determined by resource needs). The implementation technique of breaking down machines into smaller guests and their resources into smaller, sharable units, supports performance by allowing finer-grained and distributed control over resource management.

  \item \textbf{portability}: There are many dimensions to portability in computing, but in the context of multitenant infrastructures some uniquely relevant dimensions are developing guests in a standardized way---without any special knowledge of the environment where they will be deployed---and abstracting deployment and management across physical machines, limiting dependence on low-level hardware details. For example, a container guest can be deployed anywhere in the cluster, or a virtual machine guest can be deployed on any compute machine in the cloud. The implementation techniques of standardizing interfaces so guests are substitutable and hiding implementation and hardware details behind well-defined interfaces both support portability.

  \item \textbf{process}: The early literature tended to use the terms ``job'' \cite{rochester_computer_1955} or ``program'' \cite{codd_multiprogramming_1959, opler_multiprogramming:_1959, amdahl_architecture_1964}, and ``process'' only appeared around the mid-1960s \cite{dennis_programming_1966, ackerman_implementation_1967}. This survey uses the modern term ``process''. The early use of ``multiprogramming'' meaning ``multiprocessing'' was derived from the early use of ``program'' meaning ``process''.

  \item \textbf{security}: There are many dimensions to security in computing, but in the context of multitenant infrastructures some uniquely relevant dimensions are limiting access between guests, from guests to the host, and from the host to the guests. The implementation technique of isolation supports security, at both the software level and the hardware level, by reducing the likelihood of a breach and limiting the scope of damage when a breach occurs.

  \item \textbf{virtual machine}: This survey uses the term ``virtual machine'' for multitenant deployment techniques involving the replication/emulation of real hardware architectures in software  (in contrast with \textit{container}, as defined above). The code responsible for managing virtual machine guests on a physical host machine is often called a ``hypervisor'' or ``virtual machine monitor'', both derived from early terms for the kernel, ``supervisor'' and ``monitor''. In many early implementations of virtual machines, the host kernel managed both guests and ordinary processes.

\end{itemize}

\section{Common origins} \label{sec-common-origins}

The origins of both virtual machines and containers can be traced to a fundamental shift in hardware and software architectures toward the late 1950s. The hardware of the time introduced the concept of \textit{multiprogramming}, which included both basic multitasking in the form of simple context-switching and basic multiprocessing in the form of dedicated I/O processors and multiple CPUs. Codd \cite{codd_multiprogramming_1962} attributed the earliest known use of the term multiprogramming to Rochester \cite{rochester_computer_1955} in 1955, describing the ability of an IBM 705 system to interrupt an I/O process (tape read), run a process (calculation) on the data found, and then return to the I/O process. The concept of multiprogramming evolved over the remainder of the decade through work on the EDSAC \cite{wilkes_magnetic-tape_1956}, UNIVAC LARC \cite{eckert_univac-larc_1957}, STRETCH (IBM 7030) \cite{dunwell_design_1957, codd_multiprogramming_1959}, TX-2 \cite{frankovich_functional_1957}, and an influential and comprehensive review by Gill \cite{gill_parallel_1958}. Key trade-offs discussed in the literature on multiprogramming---around security, performance, portability, and complexity---continue to echo through modern literature on virtual machines and containers.

\subsection {Security}

Multiprogramming increased the complexity of the system software---due to simultaneous and interleaved processes interacting with other processes and shared hardware resources---and also increased the consequences of misbehaving system software---since any process had the potential to disrupt any other process on the same machine. Codd \textit{et al.} \cite{codd_multiprogramming_1959} discussed secure isolation as a requirement for ``noninterference'' between processes regarding errors, in the core design principles for STRETCH. Codd \cite{codd_multiprogramming_1962} later expanded on the requirement as a need to prevent processes from making ``accidental or fraudulent'' changes to another process. Buzen and Gagliardi \cite{buzen_evolution_1973} called out the risk of one process modifying memory allocated to other processes or privileged system operations.

In response to the increase in complexity and risk, system software of the time introduced a familiar form of isolation, granting a small privileged kernel of system software unrestricted access to all hardware resources and running processes, as well as responsibility for potentially disruptive operations such as memory and storage allocation, process scheduling, and interrupt handling, while restricting access to such features from any software outside the kernel. Codd \textit{et al.} \cite{codd_multiprogramming_1959} described the structure and function of the STRETCH kernel in detail, including concurrency, interrupts, memory protection, and time limits (an early form of resource usage control). Amdahl \textit{et al.} \cite{amdahl_architecture_1964} touched on the separation of the kernel in the IBM System/360, including appendices of relevant opcodes and protected storage locations. Opler and Baird \cite{opler_multiprogramming:_1959} weighed trade-offs around having the kernel take responsibility for coordinating the parallel operation of processes, and judged the approach to have potential to improve portability of programs not written for parallel operation, as well as potential to minimize complexity for programmers who would no longer be responsible to manually coordinate the parallel operation of each program.

\subsection{Performance}

One of the fundamental goals of adding multiprogramming to hardware and operating systems in the late 1950s was to improve performance through more efficient utilization of available resources by sharing them across parallel processes. Codd \textit{et al.} \cite{codd_multiprogramming_1959} described performance as a requirement for ``noninterference'' between processes regarding ``undue delay''. Opler and Baird \cite{opler_multiprogramming:_1959} explored the trade-offs between the performance advantages of increasing utilization through multiprocessing, versus the increased complexity of developing for such systems. Codd published two further papers in 1960 \cite{codd_multiprogram_1960, codd_multiprogram_1960-1} about performance considerations for process scheduling algorithms in multiprogramming. Amdahl \textit{et al.} \cite[p.~89]{amdahl_architecture_1964} explored the trade-offs between performance and portability in the architecture design of the IBM System/360. Dennis \cite[p.~590]{dennis_segmentation_1965} noted the performance advantages of dynamic memory allocation for multiprogramming.

\subsection{Portability}

In the 1950s, it was common for specialized system software to be developed for each new model of hardware, requiring programs to be rewritten to run on even closely-related machines. As the system software and programs grew larger and more complex, the porting effort grew more costly, motivating a desire for programs to be portable across different machines. Codd \textit{et al.} \cite{codd_multiprogramming_1959} discussed portability as a requirement for ``independence of preparation'' and ``flexible allocation of space and time''. Amdahl \textit{et al.} \cite[p.~97]{amdahl_architecture_1964} emphasized portability as one of the primary design goals of the IBM System/360, specifically allowing machine-language programs to run unmodified across six different hardware models, with a variety of different configurations of peripheral devices. Buzen and Gagliardi \cite{buzen_evolution_1973} noted that the introduction of a privileged kernel compounded the problem of portability, since a program might have to be rewritten to run on two different kernels, even when the underlying hardware was compatible or completely identical.

\subsection{Minimizing complexity}

Another early realization after the introduction of multiprogramming was that it was unreasonable to expect the developer of each process to directly manage all the complexity of interacting with every other process running on the machine, so the privileged kernel approach had the advantage of allowing processes to maintain a more minimal focus on their own internals. Codd \textit{et al.} \cite{codd_multiprogramming_1959} described minimizing complexity as a requirement for ``minimum information from programmer''. Nearly a decade before Rushby first wrote about the idea of a Trusted Computing Base \cite{rushby_design_1981}, Buzen and Gagliardi \cite[p.291]{buzen_evolution_1973} argued for minimizing complexity within the privileged kernel, noting that such separation was effective when the privileged code base was kept small, so it could be maintained in a relatively stable state, with limited changes over time, by a few expert developers.

\section{Early virtual machines} \label{sec-early-vms}

The early work on virtual machines grew directly out of the work on multiprogramming, continuing the goal of safely sharing the resources of a physical machine across multiple processes. Initially, the idea was no more than a refinement on memory protection between processes, but it expanded into a much bigger idea: that small isolated bundles of shared resources from the host machine could present the illusion of being a physical machine running a full operating system.

\subsection{M44/44X}

In 1964, Nelson \cite{nelson_mapping_1964} published an internal research report at IBM outlining plans for an experimental machine based on the IBM 7044, called the M44. The project built on earlier work in multiprogramming, improving process isolation and scheduling in the privileged kernel with an early form of virtual memory. They called the memory mapped for a particular process a ``virtual machine'' \cite[p.~14]{nelson_mapping_1964}. The 44X part of the name stood for the virtual machines (also based on the IBM 7044) running on top of the M44 host machine.

Nelson \cite[p.~4-6]{nelson_mapping_1964} identified the performance advantages of dynamically allocated shared resources (especially memory and CPU) as one of the primary motivators for the M44/44X experiments. Portability was another central consideration, allowing software to run unmodified across single process, multiprocess, and debugging contexts \cite[pp.~9-10]{nelson_mapping_1964}.

The M44/44X lacked almost all of the features we would associate with virtual machines today, but it played an important, though largely forgotten, part in the history of virtual machines. Denning \cite{denning_performance_1981} reflected that the M44/44X was central to significant theoretical and experimental advances in memory research around paging, segmentation, and virtual memory in the 1960s. 

\subsection{Cambridge Monitor System}

The IBM System/360 was explicitly designed for portability of software across different models and different hardware configurations \cite{amdahl_architecture_1964}. In the mid-1960s, IBM's \textit{Control Program-40 Cambridge Monitor System} (CP-40/CMS) project running on a modified IBM System/360 (model 40) took the idea a few steps further---initially calling the work a ``pseudo-machine'', but later adopting the term ``virtual machine'' \cite[p.~485]{creasy_origin_1981}. The CP-40/CMS and later CP-67/CMS\footnote{For the IBM System/360 model 67.} projects improved on earlier approaches to portability, making it possible for software written for a bare metal machine to run unmodified in a virtual machine, which could simulate the appearance of various different hardware configurations \cite[pp.~1-2]{adair_virtual_1966}. It also improved isolation by introducing privilege separation for interrupts \cite[pp.~6-7]{adair_virtual_1966}, paged memory within virtual machine guests \cite{buzen_evolution_1973, parmelee_virtual_1972}, and simulated devices \cite{buzen_evolution_1973, noauthor_control_1971}. IBM's work on the CP-40/CMS focused on improving performance through efficient utilization of shared memory \cite[pp.~3-5]{adair_virtual_1966}, and explictly did not target efficient utilization of CPU through sharing \cite[p.~1]{adair_virtual_1966}. Kogut \cite{kogut_segment_1973} developed a variant of CP-67/CMS to improve performance through dynamic allocation of storage (physical disk) to virtual machines.

\subsection{VM/370}

IBM's VM/370 running on the System/370 hardware followed in the early 1970s, and included virtual memory hardware \cite[p.~485]{creasy_origin_1981}. Madnick and Donovan \cite[p.~214]{madnick_application_1973} estimated the overhead of the VM/370 at 10-15\%, but deemed the performance trade-off to be worthwhile from a security perspective. Goldberg \cite[pp.~39-40]{goldberg_survey_1974} identified the source of overhead as primarily: maintaining state for virtual processors, trapping and emulating privileged instructions, and memory address translation for virtual machine guests (especially when paging was supported in the guests). In retrospect, Creasy noted that efficient execution was never a primary goal of IBM's work on the CP-40, CP-67, or VM/370 \cite[p.~487]{creasy_origin_1981}, and the focus was instead on efficient utilization of available resources \cite[p.~484]{creasy_origin_1981}.

\subsection{Trade-offs}

In their formal requirements for virtual machines in the mid-1970s, Popek and Goldberg \cite[p.~413]{popek_formal_1974} stated that ideally virtual machines should ``show at worst only minor decreases in speed'' compared to running on bare metal. In 2017, Bugnion \textit{et al.} \cite{bugnion_hardware_2017} explained Popek and Goldberg's requirements in modern terms, exploring the performance impact for hardware architectures that do not fully meet the requirements.

Buzen and Gagliardi \cite[p.~291]{buzen_evolution_1973}, Madnick and Donovan \cite[p.~212]{madnick_application_1973}, Goldberg \cite[p.~75]{goldberg_architecture_1973}, and Creasy \cite[p.~486]{creasy_origin_1981} all observed that the portability offered by virtual machines was also an advantage for development purposes, since it allowed development and testing of multiple different versions of the kernel/operating systems---and programs targeting those kernels/operating systems---in multiple different virtual hardware configurations, on the same physical machine at the same time.

Buzen and Gagliardi \cite{buzen_evolution_1973} considered one of the key advantages of the virtual machine approach to be that ``virtual machine monitors typically do not require a large amount of code or a high degree of logical complexity''. Popek and Kline \cite[p.~294]{popek_verifiable_1975} discussed the advantage of virtual machines being smaller and less complex than a kernel and complete operating system, improving their potential to be secure. Goldberg \cite[p.~39]{goldberg_survey_1974} suggested minimizing complexity as a way to improve performance: selectively disabling more expensive features (such as memory paging in guests) for virtual machines that would not use the features. Creasy \cite[p.~488]{creasy_origin_1981} discussed the advantages of minimizing interdependencies between virtual machines, giving preference to standard interfaces on the host machine.

A frequently-cited group of papers in the early 1970s, by Lauer and Snow \cite{lauer_is_1972}, Lauer and Wyeth \cite{lauer_recursive_1973}, and Srodawa and Bates \cite{srodawa_efficient_1973}, suggested that virtual machines offered a sufficient level of isolation that it was no longer necessary to maintain a privilege-separated kernel in the host operating system. However, by that point in time the concept of a privileged kernel was well enough established that the idea of eliminating it bordered on heresy. Buzen and Gagliardi \cite[p.~297]{buzen_evolution_1973} observed that the proposal depended heavily on the ability of the virtual machine implementation to handle all virtual memory mapping directly, but since the papers failed to take memory segmentation into account, the approach could not be implemented as initially proposed.

\subsection{Decline}

As companies like DEC, Honeywell, HP, Intel, and Xerox introduced smaller hardware to the market in the 1970s, they did not include hardware support for features such as virtual memory and the ability to trap all sensitive instructions, which made it challenging to implement strong isolation using virtual machine techniques on such hardware \cite{dickman_small_1973, galley_pdp-10_1973}. Creasy \cite[p.~484]{creasy_origin_1981} observed in the early 1980s that the advent of the personal computer decreased interest in the early forms of virtual machines---which were largely developed for the purpose of isolating users in time-sharing systems on mainframes---but he recognized potential for virtual machines to serve ``the future's network of personal computers''.\footnote{It was a reasonable prediction for the time: HTTP was introducted much later in the 1980s, but the RFC for the Internet Protocol (IP) \cite{postel_internet_1981} was published in the same month as Creasy's article, and TCP had already been around since the mid-1970s.}

\section{Early capabilities} \label{sec-capabilities}

The origin of containers is often attributed \cite{combe_docker_2016, kovacs_comparison_2017, li_performance_2017, raho_kvm_2015, bernstein_containers_2014} to the addition of the \texttt{chroot} system call in the Seventh Edition of UNIX released by Bell Labs in 1979 \cite{kernighan_unix_1979}. The simple form of filesystem namespace isolation that \texttt{chroot} provides was certainly one influence on the development of containers, though it lacked any concept of isolation for process namespaces \cite{kamp_jails:_2000, priedhorsky_charliecloud:_2017}. However, containers are not a single technology, they are a collection of technologies combined to provide secure isolation, including namespaces, cgroups, seccomp, and capabilities. Combe \textit{et al.} \cite{combe_docker_2016}, Jian and Chen \cite{jian_defense_2017}, Kov{\'a}cs \cite{kovacs_comparison_2017}, Priedhorsky and Randles \cite{priedhorsky_charliecloud:_2017}, and Raho \textit{et al.} \cite{raho_kvm_2015} describe how these different technologies combine to provide secure isolation for containers. It is more accurate to attribute the origin of containers to the earliest of these technologies, capabilities, which began decades before \texttt{chroot} and several years before the first work on virtual machines. Like containers, capabilities took the approach of building secure isolation into the hardware and the operating system, without virtualization.

\subsection{Descriptors}

In the early 1960s, inspired by the need to isolate processes, the Burroughs B5000 hardware architecture introduced an improvement to memory protection called \textit{descriptors}, which flagged whether a particular memory segment held code or data, and protected the system by ensuring it could only execute code (and not data), and could only access data appropriately (a single element scalar, or bounds-checked array) \cite{mayer_architecture_1982, levy_capability-based_1984}. A process on the B5000 could only access its own code and data segments through a private Program Reference Table, which held the descriptors for the process \cite[p.~23]{levy_capability-based_1984}. A descriptor also flagged whether a segment was actively in main memory or needed to be loaded from drum \cite[p.~24]{levy_capability-based_1984}.

\subsection{Dennis and Van Horn}

In the mid-1960s, Dennis and Van Horn \cite{dennis_programming_1966} introduced the term \textit{capability} in theoretical work directly inspired by both the Burroughs B5000 and MIT's Compatible Time-Sharing System (CTSS) \cite[p.~154]{dennis_programming_1966}. Like the B5000 descriptors, capabilities defined the set of memory segments a process was permitted to read, write, or execute \cite[p.~42]{levy_capability-based_1984}. These early capabilities introduced several important refinements: a process executed within a protected \textit{domain} with an associated capability list; multiple processes could share the same capability list; and a process could \texttt{FORK} a parallel process with the same capabilities (but no greater), or create a subprocess with a subset of its own capabilities (but no greater) \cite[pp.~42-44]{levy_capability-based_1984}. These theoretical capabilities also had a concept of ownership (by a process or a user) \cite[p.~42]{levy_capability-based_1984}, and of persistent data ``directories'' (but not files) which survived beyond the execution of a process and could be private to a user or accessible to any user \cite[pp.~44-45]{levy_capability-based_1984}.

Soon after Dennis and Van Horn published their theoretical capabilities, Ackerman and Plummer \cite{ackerman_implementation_1967} implemented some aspects of capabilities relating to resource control on a modified PDP-1 at MIT, and added a file capability in addition to the directory capability---a precursor to filesystem namespaces.

\subsection{Chicago Magic Number Machine}

In 1967, the University of Chicago launched the first attempt at designing and building a general-purpose hardware and software capability system, which they later called the Chicago Magic Number Machine\footnote{The unusual name was emblematic of the decade, from Ken Kesey's ``Magic Bus'' to the Beatles' ``Magical Mystery Tour''. At the level of physical memory, capabilities are effectively a ``magic'' number.} \cite{fabry_users_1967, fabry_preliminary_1968}. The Chicago machine pushed the concept of separation between capabilities and data further, to protect against users altering the capabilities that limited their access to memory on the system \cite[pp.~49-50]{levy_capability-based_1984}. The machine had a set of physical registers for capabilities, which were distinct from the usual set of registers for data. It also flagged whether each memory segment stored capabilities or data, and prevented processes from performing data operations like reading or writing on capability segments or capability registers. Inter-process communication also sent both a capability segment and a data segment \cite[p.~51]{levy_capability-based_1984}.

The University of Chicago project ran out of funding and was never completed, but it inspired subsequent work on CAL-TSS \cite[p.~49]{levy_capability-based_1984}.

\subsection{CAL-TSS}

In 1968, the University of California at Berkeley launched the CAL-TSS project \cite[pp.~52-57]{levy_capability-based_1984}, which aimed to produce a general-purpose capability-based operating system, to run on a Control Data Corporation 6400 model (RISC architecture) mainframe machine, without any special customization to the hardware. Like previous implementations, CAL-TSS confined a process to a domain, restricting access to hardware registers, memory, executable code, system calls to the kernel, and inter-process communication. The project introduced a concept of unique and non-reusable identifiers for objects, to protect against reuse of dangling pointers to access and modify memory that has been reallocated after being freed.

The CAL-TSS project encountered difficulties implementing the operating system as designed, and was terminated in 1971. Levy \cite[p.~57]{levy_capability-based_1984} identified the memory management features of the CDC 6400 as a particularly troublesome obstacle to the implementation. In postmortem analysis, Sturgis \cite{sturgis_postmortem_1973} and Lampson and Sturgis \cite{lampson_reflections_1976} reflected that CAL-TSS ended up being large, overly complex, and slow, and attributed this primarily to a poor match between the hardware they selected and the design of mapped address spaces, and also to their design choice of distributing privileged code for manipulating global system data across individual processes, rather than consolidating it in a privileged kernel.

\subsection{Plessey System 250}

In the early 1970s, the Plessey System 250 \cite{england_capability_1974} was a commercially successful real-time multiprocessing telephone-switch controller. It implemented capabilities for memory protection and process isolation \cite[p.~65]{levy_capability-based_1984}, and expanded capabilities into the I/O system \cite[p.~77]{levy_capability-based_1984}.

\subsection{Provably Secure Operating System}

Also in the early 1970s, the Stanford Research Institute began a project to explore the potential of formal proofs applied to a capability-based operating system design, which they called the Provably Secure Operating System (PSOS)  \cite{neumann_provably_1980}. The design was completed in 1980, but never fully formally proven, and never implemented \cite{neumann_psos_2003}.

\subsection{CAP}

In the late 1970s, the University of Cambridge's CAP machine \cite{needham_cambridge_1977, wilkes_cambridge_1979} successfully implemented capabilities as general-purpose hardware combined with a complementary operating system. The CAP introduced a refinement replacing the privileged kernel with an ordinary process, so the special control the ``root'' process had over the entire system was really just the normal ability of any process to create subprocesses and grant a subset of its own capabilities to those subprocesses \cite[pp.~80-81]{levy_capability-based_1984}.

\subsection{Object systems}

Several software offshoots of the early capability systems generalized the idea by treating processes and shared resources as typed objects with associated capabilities, including Carnegie-Mellon's Hydra \cite{wulf_hydra:_1974, wulf_hydra-c._1981} and StarOS \cite{jones_staros_1979}.

\subsection{IBM System/38}

In 1978, IBM announced plans for a capability-based hardware architecture, the System/38, which they shipped in 1980 \cite[p.~137]{levy_capability-based_1984}. Berstis \cite{berstis_security_1980} characterized the primary goal of the System/38 as improving memory protection without sacrificing performance. Houdek \cite{houdek_ibm_1981} described the implementation of capabilities as protected pointers in detail. The System/38 introduced a concept of user profiles associated with protected process domains \cite[pp.~249-250]{berstis_security_1980}, which were vaguely reminiscent of modern user namespaces, though implemented differently. User profiles allowed for revocation of capabilities, but at the cost of significantly increased complexity in the implementation \cite[pp.~155-156]{levy_capability-based_1984}.

The System/38 was succeeded by the AS/400 in the late 1980s, which removed capability-based addressing \cite[p.~119]{soltis_fortress_2001}. The AS/400 later adopted the concept of logical partitioning from the IBM System/370 \cite[pp.~1-2]{schimunek_slicing_1999}, to divide the physical resources of the host machine between multiple guests at the hardware level\footnote{Unlike virtual machines, capabilities, or containers, which divide physical resources at the software level.} \cite[pp.~240,~328]{soltis_fortress_2001}.

\subsection{Intel iAPX 432}

In 1975, Intel began designing the iAPX 432 \cite{noauthor_iapx_1981} capability-based hardware architecture, which they originally intended to be their next-generation, market-leading CPU, replacing the 8080 \cite[p.~79]{mazor_intels_2010}. The project finally shipped in 1981, but it was significantly delayed and significantly over budget \cite[p.~79]{mazor_intels_2010}.

Mazor \cite[p.~75]{mazor_intels_2010} recorded that performance was not considered as a goal in the design of the iAPX 432. Hansen \textit{et al.} \cite{hansen_performance_1982} measured the performance of the iAPX 432 against the Intel 8086, Motorola 68000, and the VAX-11/780 in 1982, with results as poor as 95 times slower on some benchmarks. Norton \cite[p.~27]{norton_hardware_2016} assessed the poor performance and unoptimized compiler offered by the iAPX 432 as the leading cause of its commercial failure. Levy \cite[p.~186]{levy_capability-based_1984} blamed the commercial failure on both poor performance and over-hyped marketing.

In a move that Mazor described as ``a crash program...to save Intel's market share'' \cite[p.~75]{mazor_intels_2010}, Intel launched a parallel project to develop the 8086 architecture (the first in a long line of x86 CPUs), which became Intel's leading product line by default, rather than by design \cite[p.~79]{mazor_intels_2010}.\footnote{In hindsight, the commercial failure of the iAPX 432 probably influenced Intel's single-minded focus on performance and disinterest in memory protection techniques in the decades that followed, which ultimately contributed to the vulnerabilities discussed in Section \ref{sec-security}.}

\subsection{Trade-offs}

The early capability systems in the 1960s and 1970s sacrificed performance for the sake of security, though Levy speculated in the mid-1980s that this was partly due to ``hardware poorly matched to the task'' \cite[p.~205]{levy_capability-based_1984}. Wilkes \cite[pp.~49-59]{wilkes_time-sharing_1968} contrasted the memory protection features of capabilities with other systems of the time, including detailed descriptions of hardware implementations.

Levy \cite[p.~205]{levy_capability-based_1984} also observed that the early capability systems significantly increased complexity for the sake of security. Patterson and S\'equin \cite{patterson_risc_1981} and Patterson and Ditzel \cite{patterson_case_1980} judged this sacrifice as a major reason the capability machines were surpassed by simpler architectures, such as RISC.

Kirk McKusick recalled that the primary reason Bill Joy ported \texttt{chroot} from UNIX into BSD in 1982 was for portability, so he could build different versions of the system in an isolated build directory \cite[p.~11]{kamp_jails:_2000}.

\subsection{Decline}

As with virtual machines, interest in the early capability systems sharply declined in the 1980s, influenced by several independent factors. Several early attempts to implement capabilities were terminated uncompleted---notably the Chicago Magic Number Machine, CAL-TSS, and the Provably Secure Operating System---contributing to a reputation that capability systems were difficult to implement and perhaps overly ambitious, despite the successful implementations that followed. The commercial failure of Intel's iAPX 432 raised further doubts on the feasibility of capability-based architectures. In 2003, Neumann and Feiertag \cite[p.~6]{neumann_psos_2003} looked back on the early capability systems, expressing disappointment that ``the demand for meaningfully secure systems has remained surprisingly small until recently''.

Perhaps the most significant factor in the decline of capabilities was the rise of the general-purpose operating system, which was a third important technology that evolved from multiprogramming. MIT's Compatible Time-Sharing System (CTSS) \cite{corbato_experimental_1962, wilkes_time-sharing_1968} laid the foundation for Multics \cite{corbato_multics:_1972}, which later inspired UNIX \cite{ritchie_evolution_1980} and its robust mutation, the Berkeley Software Distribution (BSD)\footnote{One noteworthy connection between these factors is Robert Fabry, who worked on the Chicago Magic Number Machine in the 1960s \cite{fabry_users_1967, fabry_preliminary_1968} while doing a PhD at the University of Chicago \cite{fabry_list-structured_1971}, and was also the catalyst for Berkeley's interest in UNIX and substantial investment in the BSD project, while he was a professor at Berkeley in the 1970s \cite{mckusick_twenty_1999}.} \cite{mckusick_twenty_1999, mckusick_current_1989}. Saltzer and Schroeder \cite[p.~1294]{saltzer_protection_1975} contrasted capabilities with the access control list models adopted by Multics and its descendants, calling out revocation of access as one major area where capabilities fell short.

While none of the early capability systems remain in use today, they have not been entirely forgotten. In 2003, Miller \textit{et al.} \cite{miller_capability_2003} reviewed capability systems from a historical perspective, addressing common misconceptions about capabilities related to revocation, confinement, and equivalence to access control lists. Section \ref{sec-containers} traces the evolution of a feature called capabilities in the modern Linux Kernel. FreeBSD took a different approach for the feature it calls capabilities, and integrated the Capsicum framework \cite[p.~30]{mckusick_design_2014}, which was more directly derived from the classic capability systems \cite{watson_capsicum:_2010, anderson_towards_2017}. In 2012, the CHERI project \cite{watson_cheri:_2012, watson_capability_2014, woodruff_cheri_2014, watson_cheri:_2015} expanded on the ideas of the Capsicum framework, pushing its capability model down into a RISC-based hardware architecture. Since 2016, Google has been exploring a revival of capability systems with the Fuchsia operating system and Zircon microkernel \cite{google_fuchsia_2018}. In a 2018 plenary session about Spectre/Meltdown, Hennessy \cite{hennessy_era_2018} pointed to future potential for capabilities, reflecting that the early capability systems ``probably weren't the right match for what software designers thought they needed and they were too inefficient at the time'', but suggested ``those are all things we know how to fix now...so it's time, I think, to begin re-examining some of those more sophisticated [protection] mechanisms and see if they'll work''.

\section{Modern virtual machines} \label{sec-modern-vms}

Virtual machines still existed in the 1980s and 1990s, but garnered only a bare minimum of activity and interest. DOS, OS/2, and Windows all offered a limited form of DOS virtual machines during that time, though it might be more fair to categorize those as emulation. The rise of programming languages like Smalltalk and Java re-purposing the term ``virtual machine''---to refer to an abstraction layer of a language runtime, rather than a software replication of a real hardware architecture---may be indicative of how dead the original concept of virtual machines was in that period.

After a hiatus lasting nearly two decades, the late 1990s brought a resurgence of interest in virtual machines, but for a new purpose adapted to the technology of the time.

\subsection{Disco}

In 1997, the Disco research project at Stanford University explored reviving virtual machines as an approach to making efficient use of hardware with multiple CPUs (on the order of ``tens to hundreds''), and included a lightweight library operating system for guests (SPLASHOS) as an option, in addition to supporting commodity operating systems as guests. Bugnion \textit{et al.} \cite{bugnion_disco:_1997} cited portability (rather than security or performance) as the primary motivation of the Disco project, which proposed virtual machines as a potential way to allow commodity operating systems (Unix, Windows NT, and Linux) to run on NUMA architectures without extensive modifications.

\subsection{VMware}

A year later, the team behind Disco founded VMware to continue their work, and released a workstation product in 1999 \cite{bugnion_bringing_2012}, quickly followed by two server products (GSX and ESX) in 2001 \cite{waldspurger_memory_2002, ahmad_analysis_2003, sapuntzakis_optimizing_2002}. VMware faced a challenge in virtualizing the x86 architectures of the time, because the hardware did not support traditional virtualization techniques---specifically the architecture contained some sensitive instructions which were not also privileged---so a virtual machine monitor could not rely on trapping protection exceptions as the sole means of identifying when to execute emulated instructions as a safe replacement, since some potentially harmful instructions would never be trapped  \cite[p.131]{robin_analysis_2000}.\footnote{Popek and Goldberg \cite{popek_formal_1974} classically defined such machines as unvirtualizable.} To work around this limitation, VMware combined the trap-and-execute technique with a dynamic binary translation technique \cite[p.12:3]{bugnion_bringing_2012}, which was faster than full emulation, but still allowed the guest operating system to run unmodified \cite[p.12:29-36]{bugnion_bringing_2012}.

\subsection{Denali}

The Denali project at the University of Washington in 2002 \cite{whitaker_denali:_2002} introduced the term ``paravirtualization'',\footnote{The term was new, but the technique had roots stretching back to IBM's VM/370 \cite{creasy_origin_1981, goldberg_survey_1974}.} another work-around for the lack of hardware virtualization support in x86, which involved altering the instruction set in the virtualized hardware architecture, and then porting the guest operating system to run on the altered instruction set \cite{whitaker_denali:_2002-1}.

\subsection{Xen}

The Xen project at the University of Cambridge in 2003 \cite{barham_xen_2003} also used paravirtualization techniques and modified guest operating systems, but emphasized the importance of preserving the application binary interface (ABI) within the guests so that guest applications could run unmodified. Xen's greatest technical contribution may have been its approach to precise accounting for resource usage, with the explicit intention to individually bill tenants sharing physical machines \cite[p.176]{barham_xen_2003}, which was a relatively radical idea at the time,\footnote{Partially inspired by earlier work, involving some of the same authors, on resource management in the Nemesis operating system \cite{barham_fresh_1997}.} and directly led to the creation of Amazon's Elastic Compute Cloud (EC2) a couple of years later \cite{barr_amazon_2006}.\footnote{The EC2 beta was launched in 2006, but when I presented at the Amazon Developers Conference in 2005, they were already working on it.}

Chisnall \cite{chisnall_definitive_2007} provided a detailed account of Xen's architecture and design goals. Xen's approach to the problem of untrapped x86 privileged instructions was to substitute a set of \textit{hypercalls} for unsafe system calls \cite[pp.10-13]{chisnall_definitive_2007}. Smith and Nair \cite[p.422]{smith_architecture_2005} highlighted that Xen was able to run unmodified application binaries within the guest, because it ran the guest in ring 1 of the IA-32 privilege levels and the hypervisor in ring 0, so all privileged instructions were filtered through the hypervisor.

\subsection{x86 Hardware virtualization extensions}

In 2000, Robin and Irvine \cite{robin_analysis_2000} analyzed the limitations of the x86 architecture as a host for virtual machine implementations, with reference to Goldberg's earlier work \cite{goldberg_architectural_1972} on the architectural features required to support virtual machines. In the mid-2000s, in response to the growing success of virtual machines, and the challenges of implementing them on x86 hardware, Intel and AMD both added hardware support for virtualization in the form of a less privileged execution mode to execute code for the virtual machine guest directly, but selectively trap sensitive instructions, eliminating the need for binary translation or paravirtualization. Rosenblum and Garfinkel \cite{rosenblum_virtual_2005} discussed the motivations behind the added hardware support for virtualization in x86, before the changes were released. Pearce \textit{et al.} \cite[p.~7]{pearce_virtualization:_2013} contrasted binary translation, paravirtualization, and the features x86 added for hardware-assisted virtualization, clarifying the x86 virtualization extensions were not full virtualization. Adams and Agesen \cite{adams_comparison_2006} recounted the difficulties VMware encountered while integrating the x86 hardware virtualization extensions, and concluded that the new features offered no performance advantage over binary translation.

In 2007, the KVM subsystem for the Linux Kernel provided an API for accessing the x86 hardware virtualization extensions \cite{kivity_kvm:_2007}. Since KVM was only a Kernel subsystem, the developers released a fork of QEMU\footnote{Which was previously only an emulator \cite{bellard_qemu_2005}.} as the userspace counterpart of KVM, so the combination of QEMU+KVM provided a full virtual machine implementation, including virtual devices \cite[pp.128-129]{wang_isolating_2012}. Eventually, KVM support was merged into mainline QEMU \cite{liguori_qemu_2012}.

\subsection{Hyper-V}

In 2008, Microsoft released a beta of Hyper-V \cite{kappel_microsoft_2009} for Windows Server. It was built on top of the x86 hardware virtualization extensions, and for some virtual devices offered a choice between slower emulation and faster paravirtualization if the guest operating system installed the ``Enlightened I/O'' extensions. Like Xen's Dom0, Hyper-V granted special privileges to one guest, called the ``parent partition'', which hosted the virtual devices and handled requests from the other guests. 

In 2010, Bolte \textit{et al.} \cite{bolte_non-intrusive_2010} incorporated support for Hyper-V into \texttt{libvirt}, so it could be managed through a standardized interface, together with Xen, QEMU+KVM, and VMware ESX.

\subsection{Trade-offs}

Denali and Xen both used paravirtualization techniques, sacrificing portability to gain performance, but their goals for scale were completely different: Denali considered 10,000 virtual machines\footnote{On a 1.7GHz Pentium 4 with 1GB RAM.} to be a good result \cite{whitaker_scale_2002}---achieved through a combination of lightweight guests and a minimal host---while Xen argued that 100 virtual machines running full operating systems\footnote{On a 2.4GHz dual-core Xeon with 2GB RAM.} was a more reasonable target \cite[p.165,175]{barham_xen_2003}. To some extent, Denali was more in line with modern container implementations than with the virtual machine implementations of its day. Xen has shifted their estimation of required scale upward over the years, but still exhibits a tolerance for unnecessarily mediocre performance. For example, Manco \textit{et al.} \cite{manco_my_2017} demonstrated that a few small internal changes to the way Xen stores metadata and creates virtual devices improved virtual machine instantiation time by an order of magnitude---a result 50-200 times faster than Docker's container instantiation---however those patches are unlikely to ever make it into mainline Xen.

Xen and KVM have a reputation for sacrificing performance to gain security, however several independent lines of research have raised questions as to whether those security gains are real or imagined. Perez-Botero \textit{et al.} \cite{perez-botero_characterizing_2013} analyzed security vulnerabilities in Xen and KVM between 2008-2012, categorizing them by source, vector, and target, and observed that the most common vector of attack was device emulation (Xen 34\%, KVM 40\%), the majority were triggered from within the virtual machine guest (Xen 71\%, KVM 66\%), and the majority successfully targeted the hypervisor's Ring -1 privileges or slightly less privileged control over Dom0 or the host operating system (Xen 80\%, KVM 76\%). Chandramouli \textit{et al.} \cite{chandramouli_methodology_2018} built on the work of Perez-Botero \textit {et al.} \cite{perez-botero_characterizing_2013}, moving toward a more general framework for forensic analysis of vulnerabilities in virtual machine implementations. Ishiguro and Kono \cite{ishiguro_hardening_2018} evaluated vulnerabilities in Xen and KVM related to instruction emulation between 2009-2017. They demonstrated that a prototype ``instruction firewall'' on KVM---which denies emulation of all instructions except the small subset deemed legitimate in the current execution context---could have defended against the known instruction emulation vulnerabilities, however the patches are unlikely to ever make it into mainline KVM.

Szefer \textit{et al.} \cite{szefer_eliminating_2011} demonstrated in the NoHype implementation (based on Xen) that eliminating the hypervisor and running virtual machines with more direct access to the hardware improved security by reducing the attack surface and removing virtual machine exit events as potential attack vectors. However, the approach involved a performance trade-off in resource utilization that was not viable for most real deployments: it pre-allocated processor cores, memory, and I/O devices dedicated to specific virtual machines, rather than allowing for oversubscription and dynamic allocation in response to load.

One persistent argument in favor of virtual machines has been that virtual machine implementations have fewer lines of code than a kernel or host operating system, and are therefore easier to code-review and secure \cite{bugnion_disco:_1997, garfinkel_virtual_2003, manco_my_2017, pearce_virtualization:_2013, seshadri_secvisor:_2007}, which is the classic trade-off of minimizing complexity to gain security. However, less code offers only a vague potential for security, and even that potential becomes questionable as modern virtual machine implementations have grown larger and more complex \cite{colp_breaking_2011, pearce_virtualization:_2013, williams_say_2018, bratus_vm-based_2010}.

Recent work on virtual machines---such as \texttt{ukvm} \cite{williams_unikernel_2016}, LightVM \cite{manco_my_2017}, and Kata Containers (formerly Intel Clear Containers) \cite{noauthor_kata_2018}---has shifted back toward an emphasis on improving performance. However, this work appears to be founded on the assumption that the virtual machine implementations under discussion are adequately secure, and need only improve performance, which is a dubious assumption at best.

Two notable departures from this complacent attitude to security are Google's crosvm \cite{google_chrome_2018} and Amazon's Firecracker \cite{amazon_firecracker_2019}, which aim to improve both performance and security, by replacing QEMU with a radically smaller and simpler userspace component for KVM, and by choosing Rust as the implementation language for memory safety.\footnote{The memory safety features of Rust do not address the security vulnerabilities discussed in Section \ref{sec-security}, but can eliminate another common class of memory access vulnerabilities, such as buffer overflows/underflows and use-after-free. Szekeres \textit{et al.} \cite{szekeres_sok:_2013} provide a systematic account of such vulnerabilities and their impact in the C/C++ programming languages.} Firecracker started as a fork of crosvm, but the two projects are collaborating on generalizing the divergence into a set of Rust libraries they can share. 

\subsection{Decline}

Toward the end of the 2000s, the enthusiasm for virtual machines gave way to a growing skepticism. Garfinkel \textit{et al.} \cite{garfinkel_compatibility_2007} demonstrated that virtual machine environments could reliably be detected on close inspection,  reviving the long-running tension between the ideals of strong isolation in virtual machines, and the reality of actual implementations. Buzen and Gagliardi \cite{buzen_evolution_1973} commented on the ideals in the early 1970s, ``Since a privileged software nucleus has, in principle, no way of determining whether it is running on a virtual or a real machine, it has no way of spying on or altering any other virtual machine that may be coexisting with it in the same system.'' but in the same paper acknowledged, ``In practice no virtual machine is completely equivalent to its real machine counterpart.''

In 2010, Bratus \textit{et al.} \cite{bratus_vm-based_2010} criticized the myopic focus of systems security research on virtual machines and the resulting neglect of other potentially superior approaches to system security. Vasudevan \textit{et al.} \cite{vasudevan_requirements_2010} outlined a set of requirements for protecting the integrity of virtual machines implemented on x86 with hardware virtualization support, and evaluated all existing implementations as ``unsuitable for use with highly sensitive applications'' \cite[p.141]{vasudevan_requirements_2010}. Colp \textit{et al.} \cite{colp_breaking_2011} observed that multitenant environments presented new risks for virtual machine implementations, because they required stronger isolation between guests sharing the same host than was necessary when a single tenant owned the entire physical machine.

Virtual machines such as Xen, QEMU+KVM, Hyper-V, and VMware are still in active use today, but in recent years they have entirely ceded their reputation as the ``hot new thing'' to containers.

\section{Modern containers} \label{sec-containers}

The collection of technologies that make up modern container implementations started coming together years before anyone used the term ``container''. The two decade span surrounding the development of containers corresponded to a major shift in the way information about technological advances was broadcast and consumed. Exploring the socio-economic factors driving this shift is outside the scope of this survey, however, it is worth noting that the academic literature on more recent projects such as Docker and Kubernetes is largely written by outsiders providing external commentary, rather than by the primary developers of the technologies. As a result, recent academic publications on containers tend to lack the depth of perspective and insight that was common to earlier publications on virtual machines, capabilities, and security in the Linux Kernel. The dialog driving innovation and improvements to the technology has not disappeared, but it has moved away from the academic literature and into other communication channels.

\subsection{POSIX capabilities}

In the mid-1990s, the security working group of the POSIX standards project began drafting an extension to the POSIX.1 standard, called POSIX 1003.1e \cite{noauthor_protection_1997, eisfeldt_posix:_1997, grunbacher_posix_2003}, which added a feature called ``capabilities''. The implementation details of POSIX capabilities were entirely different than the early capability systems \cite[p.97]{watson_taste_2012}, but had similarities on a conceptual level: POSIX capabilities were a set of flags associated with a process or file, which determined whether a process was permitted to perform certain actions; a process could \texttt{exec} a subprocess with a subset of its own capabilities; and the specification attempted to support the principle of least privilege \cite{noauthor_protection_1997}. However, the POSIX capabilities did not adopt the concepts of small access domains and no-privilege defaults, which were crucial elements of secure isolation in the early capability systems \cite{denning_fault_1976}. The POSIX.1e draft was withdrawn from the process in 1998 and never formally adopted as a standard \cite[p.259]{grunbacher_posix_2003}, but it formed the basis of the capabilities feature added to the Linux Kernel in 1999 (release 2.2) \cite{noauthor_capabilities7_2018, margery_capabilities_2004}.

\subsection{Namespaces and resource controls}

A second important strand in the evolution of modern container implementations was the isolation of processes via namespaces and resource usage controls. In 2000, FreeBSD added Jails \cite{kamp_jails:_2000}, which isolated filesystem namespaces (using \texttt{chroot}), but also isolated processes and network resources, in such a way that a process might be granted root privileges inside the jail, but blocked from performing operations that would affect anything outside the jail. In 2001, Linux VServer \cite{soltesz_container-based_2007} patched the Linux Kernel to add resource usage limits and isolation for filesystems, network addresses, and memory. Around the same time, Virtuozzo (later released as OpenVZ) \cite{huang_efficiently_2008, matthews_quantifying_2007} also patched the Linux Kernel to add resource usage limits and isolation for filesystems, processes, users, devices, and interprocess communication (IPC). In 2003, Nagar \textit{et al.} \cite{nagar_class-based_2003} proposed a framework for resource usage control and metering called Class-based Kernel Resource Management (CKRM), and later released it as a set of patches to the Linux Kernel.

In 2002, the Linux Kernel (release 2.4.19) introduced a filesystem namespaces feature \cite{kerrisk_namespaces_2013}.\footnote{Partially inspired by the namespaces feature of Plan 9 \cite{pike_use_1993} from Bell Labs.} In 2006, Biederman \cite{biederman_multiple_2006} proposed expanding the idea of namespace isolation in the Linux Kernel beyond the filesystem to process IDs, IPC, the network stack, and user IDs. The Kernel developers accepted the idea, and the patches to implement the features landed in the Kernel between 2006 and 2013 (releases 2.6.19 to 3.8) \cite{kerrisk_namespaces_2013}. The last set of patches to be completed was user namespaces, which allow an unprivileged user to create a namespace and grant a process full privileges for operations inside that namespace, while granting it no privileges for operations outside that namespace \cite{noauthor_user_namespaces7_2018}. The way user namespaces are nested bears a resemblance to Dennis and Van Horn's \cite{dennis_programming_1966} capabilities, where processes created more restricted subprocesses.

In 2004, Solaris added Zones \cite{price_solaris_2004} (sometimes also called Solaris Containers), which isolated processes into groups that could only observe or signal other processes in the same group, associated each zone with an isolated filesystem namespace, and set limits for shared resource consumption (initially only CPU). Between 2006 and 2007, Rohit Seth and Paul Menage worked on a patch for the Linux Kernel for a feature they called ``process containers'' \cite{corbet_process_2007}---later renamed to \textit{cgroups} for ``control groups''---which provided resource limiting, prioritization, accounting,\footnote{Similar in idea, though not in implementation, to Xen's resource usage accounting.} and control features for processes.

\subsection{Access control and system call filtering}

A third set of relevant features in the Linux Kernel evolved around secure isolation of processes through restricted access to system calls. In 2000, Cowan \textit{et al.} \cite{cowan_subdomain:_2000} released SubDomain, a Linux Kernel module which added access control checks to a limited set of system calls related to executing processes. In 2001, Loscocco and Smalley \cite{loscocco_integrating_2001} published an architectural description of SELinux, which implemented mandatory access control (MAC) for the Linux Kernel. The access control architecture of SELinux was received positively, but the implementation was rejected for being too tightly coupled with the kernel. So, in 2002, Wright \textit{et al.} \cite{wright_linux_2002} proposed the Linux Security Modules (LSM) framework as a more general approach to extensible security in the Linux Kernel, which made it possible for security policies to be loaded as Kernel modules. LSM is not an access control mechanism, but it provides a set of hooks where other security extensions such as SELinux or AppArmor can insert access control checks. LSM and a modified version of SELinux based on LSM were both merged into the mainline Linux Kernel in 2003. In 2004-2005, SubDomain was rewritten to use LSM, and rebranded under the name AppArmor.

In 2005, Andrea Arcangeli \cite{arcangeli_[patch]_2005} released a set of patches to the Linux Kernel called \textit{seccomp} for ``secure computing'', which restricted a process so that it could only run an extremely limited set of system calls to exit/return or interact with already open filehandles, and terminated a process attempting to run any other system calls. The patches were merged into the mainline Kernel later that year. However, the features of the original seccomp were inadequate and rarely used, and over the years multiple proposals to improve seccomp were unsuccessful. Then, in 2012, Will Drewry \cite{drewry_dynamic_2012} extended seccomp to allow filters for system calls to be dynamically defined using Berkeley Packet Filter (BPF) rules, which provided enough flexibility to make seccomp useful as an isolation technique. In 2013, Krude and Meyer \cite{krude_versatile_2013} implemented a framework for isolating untrusted workloads on multitenant infrastructures using seccomp system call filter policies written in BPF.

\subsection{Cluster management}

A fourth relevant strand of technology evolved around resource sharing in large-scale cluster management. In 2001, Lottiaux and Morin \cite{lottiaux_containers:_2001} used the term ``container'' for a form of shared, distributed memory which provided the illusion that multiple nodes in an SMP cluster were sharing kernel resources, including memory, disk, and network. In 2002, the Zap project \cite{osman_design_2002} used the term ``pod''\footnote{Given as an acronym for a \textbf{P}r\textbf{O}cess \textbf{D}omain abstraction.} for a group of processes sharing a private namespace, which had an isolated view of system resources such as process identifiers and network addresses. These pods were self-contained, so they could be migrated as a unit between physical machines. In the mid-2000s, Google deployed a cluster management solution called Borg \cite{verma_large-scale_2015, burns_borg_2016} into production, to orchestrate the deployment of their vast suite of web applications and services. While the code for Borg has never been seen outside Google, it was the direct inspiration for the Kubernetes project a decade later \cite[p.18:13-14]{verma_large-scale_2015}---the Borg \textit{alloc} became the Kubernetes \textit{pod}, Borglets became Kubelets, and tasks gave way to containers. Burns \textit{et al.} \cite[p.70]{burns_borg_2016} explained that improving performance through resource utilization was one of the primary motivations for Borg.

\subsection{Combined features}

The strength of modern containers is not in any one feature, but in the combination of multiple features for resource control and isolation. In 2008, Linux Containers (LXC) \cite{noauthor_linux_2018} combined cgroups, namespaces, and capabilities from the Linux Kernel into a tool for building and launching low-level system containers. Miller and Chen \cite{miller_exercise_2012} demonstrated that filesystem isolation between LXC containers could be improved by applying SELinux policies. Xavier \textit{et al.} \cite{xavier_performance_2013} and Raho \textit{et al.} \cite{raho_kvm_2015} contrasted LXC's approach to isolation and resource control using standard Linux Kernel features such as cgroups and filesystem, process, IPC, and network namespaces, versus the approaches taken by Linux VServer and OpenVZ using custom patches to the Linux Kernel to provide similar features.

Docker \cite{merkel_docker:_2014} launched in 2013 as a container management platform built on LXC. In 2014, Docker replaced LXC with \texttt{libcontainer}, its own implementation for creating containers, which also used Linux Kernel namespaces, cgroups, and capabilities \cite{hykes_docker_2014, raho_kvm_2015}. Morabito \textit{et al.} \cite{morabito_hypervisors_2015} compared the performance of LXC and Docker after the transition to libcontainer, and found them to be roughly equivalent on CPU performance, disk I/O, and network I/O, however LXC performed 30\% better on random writes, which may have been related to Docker's use of a union file system. Raho \textit{et al.} \cite{raho_kvm_2015} contrasted the implementations of Docker, QEMU+KVM, and Xen on the ARM hardware architecture. Mattetti \textit{et al.} \cite{mattetti_securing_2015} experimented with dynamically generating AppArmor rules for Docker containers based on the application workload they contained. Catuogno and Galdi \cite{catuogno_evaluation_2016} performed a case study of Docker using two different models for security assessment. They built on the work of Reshetova \textit{et al.} \cite{reshetova_security_2014} in classifying vulnerabilities by the goal of the attack: denial of service, container compromise, or privilege escalation.

In 2015, Docker split the container runtime out into a separate project, \texttt{runc}, in support of a vendor-neutral container runtime specification maintained by the Open Container Initiative (OCI). Hykes \cite{hykes_introducing_2015} highlighted that SELinux, AppArmor, and seccomp were all standard supported features in \texttt{runc}.  Koller and Williams \cite{koller_will_2017} observed that \texttt{runc} was more minimal than the Docker runtime, while still using the same isolation mechanisms from the Linux Kernel, such as namespaces and cgroups. In 2016, Docker and CoreOS merged their container image formats into a vendor-neutral container image format specification, also at OCI \cite{boulle_celebrating_2016}.

\subsection{Orchestration}

In 2014, Docker began working on Swarm, described as a clustering system for Docker, which they ultimately released late in 2015 \cite{luzzardi_announcing_2015}. Also in 2014, Google began developing Kubernetes, an orchestration tool for deploying and managing the lifecycle of containers, which they released in the middle of 2015 \cite{brewer_kubernetes_2015}. Also in 2014, Canonical began developing LXD, a container orchestration tool for LXC containers, which they released in 2016 \cite{graber_lxd_2016}.

Verma \textit{et al.} \cite{verma_large-scale_2015} outlined the design goals behind Kubernetes, in the context of lessons learned from Borg. Syed and Fernandez \cite {syed_container_2017, syed_reference_2018} pointed out that the performance advantages of the higher-level container orchestration tools, such as Kubernetes and Docker Swarm, were primarily a matter of improving resource utilization. They also contrasted the portability advantages of managing containers across multiple physical host machines against the increased complexity required for the orchestration tools to advance beyond managing a single machine host. Souppaya \textit{et al.} \cite{souppaya_application_2017} systematically reviewed increased security risks and mitigation techniques for container orchestration tools. Bila \textit{et al.} \cite{bila_leveraging_2017} extended Kubernetes with a vulnerability scanning service and network quarantine for containers.

\subsection{Trade-offs}

Containers have a reputation for substantially better performance than virtual machines, however that reputation may not be deserved. In 2015, Felter \textit{et al.} \cite{felter_updated_2015} measured the performance of Docker against QEMU+KVM and determined that neither had significant overhead on CPU and memory usage, but that KVM had a 40\% higher overhead in I/O. They observed that the overhead was primarily due to extra cycles on each I/O operation, so the impact could be mitigated for some applications by batching multiple small I/O operations into fewer large I/O operations. In 2017, Kov{\'a}cs \cite{kovacs_comparison_2017} compared CPU execution time and network throughput between Docker, LXC, Singularity, KVM, and bare metal and determined that there was no significant variation between them, as long as Docker and LXC were running in host networking mode, but in Linux bridge mode Docker and LXC exhibited high retransmission rates that negatively impacted their throughput compared to the others. Manco \textit{et al.} \cite{manco_my_2017} demonstrated that Xen virtual machine instantiation could be 50-200 times faster than Docker container instantiation, with a few low-level modifications to Xen's control stack.

Secure isolation technologies have been the core of modern container implementations from the beginning, so it would be reasonable to expect that containers would provide a strong form of isolation. However, early implementations of containers were prone to preventable security vulnerabilities, which may indicate that security was not a primary design consideration, at least not initially. Combe \textit{et al.} \cite{combe_docker_2016} analyzed security vulnerabilities in Docker and \texttt{libcontainer} between 2014-2015, and determined that the majority were related to filesystem isolation, which led to privilege escalation when Docker was run as the root user. They also suggested that some of Docker's sane default configurations for the isolation features of the Linux Kernel could be easily switched to less secure configurations through standard options to the \texttt{docker} command-line tool or the Docker daemon, and so might be prone to user error. Martin \textit{et al.} \cite{martin_docker_2018} surveyed vulnerabilities in Docker images, \texttt{libcontainer}, the Docker daemon, and orchestration tools, as well as the unique security challenges of containers in multitenant infrastructures. In addition to security patches for specific privilege escalation vulnerabilities, there has been ongoing work to integrate support for user namespaces into Docker and Kubernetes,\footnote{Such as Suda and Scrivano \cite{suda_rootless_2019} and Suda \cite{suda_allow_2019}.} so they can run as a non-root user and limit the scope of damage from privilege escalation. However, the user namespaces feature itself has had a series of vulnerabilities\footnote{Such as CVE-2018-6559, CVE-2018-18955, CVE-2014-9717, and CVE-2014-4014.} related to interfaces in the Kernel that were written with the expectation of being restricted to the root user, but are now exposed to unprivileged users.

One significant difference between virtual machine implementations and container implementations is that containers share a kernel with the host operating system, so efforts to secure the kernel greatly impact the security of containers. Reshetova \textit{et al.} \cite{reshetova_security_2014} considered the set of secure isolation features offered by the Linux Kernel as of 2014 (in the context of LXC), and judged them to have caught up with the features of FreeBSD Jails and Solaris Zones, but highlighted some areas for improvement in support of containers. These improvements included integrating Mandatory Access Control (MAC) into the Kernel as ``security namespaces''; providing a way to lock down device hotplug features for containers; and extending cgroups to support all resource management features supported by rlimits. Gao \textit{et al.} \cite{gao_containerleaks:_2017} discussed the risks of certain types of information that containers can currently access from the Linux Kernel via \textit{procfs} and \textit{sysfs}---which can be exploited to detect co-resident containers and precisely target power consumption spikes to overload servers---and prototyped a power-based namespace to partition the information for containers.

Some more recent approaches to secure isolation for containers have been inspired by virtual machine implementations. Kata Containers (formerly Intel Clear Containers) \cite{noauthor_kata_2018} wraps each Docker container or Kubernetes pod in a QEMU+KVM virtual machine \cite{noauthor_kata_2019}. They realized that QEMU was not ideal for the purpose---since it introduces a substantial performance hit compared to running bare containers, and the majority of the code relates to emulation which is not useful for wrapping containers---so a group at Intel started working on a stripped-down version of QEMU called NEMU \cite{noauthor_nemu_2018}. X-Containers \cite{shen_x-containers:_2019} used Xen's paravirtualization features to improve isolation between containers and the host, but made an unfortunate trade-off of removing isolation between containers running on the same host. Nabla Containers \cite{noauthor_nabla_2018} and gVisor \cite{google_gvisor_2019} have both taken an approach of improving isolation by heavily filtering system calls from containers to the host kernel, which is a common technique for modern virtual machines. 

Bratus \textit{et al.} \cite{bratus_vm-based_2010} noted that the ``self-protection'' techniques employed by container implementations are a necessary path for future research, since even virtual machines depend on those techniques to protect themselves. Hosseinzadeh \textit{et al.} \cite{hosseinzadeh_security_2016} explored the possibility that container implementations might directly adapt earlier work (primarily Berger \textit{et al.} \cite{berger_vtpm:_2006}) for virtual machine implementations to integrate a Trusted Platform Module (TPM) as a virtual device.

Container implementations have a potential advantage over virtual machine implementations in addressing the problem of secure isolation over the long-term, not because any existing implementations are inherently superior, but because containers take a modular approach to implementation that permits them to be more flexible over time and across different underlying software\footnote{Such as \texttt{pledge} and \texttt{unveil} on OpenBSD versus capabilities and namespaces on Linux.} and hardware architectures, as new ideas for secure isolation evolve.

\section{Security outlook} \label{sec-security}

A series of vulnerabilities related to speculative execution and side-channel attacks rose to attention over the past year. These vulnerabilities collectively upend traditional notions of secure isolation. The current reactionary approach---patching up each vulnerability as it is revealed---works in the short-term, but is a losing battle in the long-term.\footnote{Metaphorically reminiscent of the proverbial small Dutch child attempting to protect the village from flooding by inserting a tiny finger in each leak that springs in the floodbank wall.}

Early in 2018, Kocher \textit{et al.} \cite{kocher_spectre_2018} and Lipp \textit{et al.} \cite{lipp_meltdown_2018} published a set of vulnerabilities, respectively called Spectre and Meltdown, using techniques involving speculative execution and out-of-order execution. Spectre affects Intel, AMD, and ARM \cite[p.3]{kocher_spectre_2018}, can be launched from any user process (including JavaScript code run in a browser) \cite[p.3]{kocher_spectre_2018}, and grants access to any memory an attacked process could normally access \cite[p.5]{kocher_spectre_2018}. Meltdown affects Intel x86 architecture, can be launched from any user process, and grants full access to any physical memory on the same machine including kernel memory and memory allocated to any other process \cite[p.1]{lipp_meltdown_2018}. In July 2018, Schwarz \textit{et al.} \cite{schwarz_netspectre:_2018} published a remote variant of Spectre, nicknamed NetSpectre, which is launched through packets over the network and grants access to any physical memory accessible to the attacked process. In August 2018, Van Bulck \textit{et al.} \cite{van_bulck_foreshadow:_2018} published a variant of Meltdown, nicknamed Foreshadow or more broadly ``L1 Terminal Fault'' (L1TF), which is launched from unprivileged user space, and grants access to the L1 data cache, including encrypted data from Intel's Software Guard eXtensions (SGX). In November 2018, Canella \textit{et al.} \cite{canella_systematic_2018} reviewed the broad range of speculative execution vulnerabilities and proposed a comprehensive classification of the known variants and mitigations, which also revealed several previously unknown variants.

The models of secure isolation employed by virtual machines and containers offer little protection from the speculative execution vulnerabilities. Containers are vulnerable to Meltdown, though virtual machines are not because they run a different kernel than the host \cite[p.12]{lipp_meltdown_2018}. Both virtual machines and containers are vulnerable to Spectre \cite[p.3,5,6]{noauthor_speculative_2018}, NetSpectre \cite[p.11]{schwarz_netspectre:_2018}, and L1TF \cite{weisse_foreshadow-ng:_2018}, with varying degrees of compromise. Variants of L1TF\footnote{Notably CVE-2018-3646.} are especially troublesome for virtual machines, because they allow an unprivileged process in the user space of a guest to access any memory on the physical machine, including memory allocated to other guests, the host operating system, and host kernel \cite{noauthor_l1tf_2019}. Multitenant infrastructures generally allow any tenant to deploy a virtual machine or container on any physical machine in the cloud or cluster, which means it is viable to exploit these vulnerabilities by simply creating an account with a public provider and deploying malicious guests repeatedly, until one of them lands on a physical host with interesting secrets to steal.

The techniques behind the speculative execution vulnerabilities were not new, but the combined application of the techniques was more sophisticated, and the security impact more severe, than previously considered possible. Although these vulnerabilities were only recently discovered and published by defensive security researchers,\footnote{Also known as ``white hat hackers''.} it is possible that offensive security researchers\footnote{Also known as ``black hat hackers''.} discovered and exploited them much earlier, and continue to exploit additional unpublished variants. While mitigation patches have typically been applied quickly for the known variants of these vulnerabilities \cite{noauthor_speculative_2018, noauthor_retpoline:_2018}, it is not feasible to entirely disable speculative execution \cite[p.11]{kocher_spectre_2018} and out-of-order execution \cite[p.14]{lipp_meltdown_2018}, which are the primary vectors of the attacks, because the performance penalty is prohibitive, and in some cases the hardware simply has no mechanism to disable the features. The probability of further variants being discovered in the coming years is high. A substantial rethink of the fundamental hardware architecture could potentially eliminate the entire class of vulnerabilities, but in the research, development, and production timelines common to hardware vendors such a significant change could take decades.

Two notable alternative hardware architectures, CHERI and RISC-V, were already under development before the flood of speculative execution vulnerabilities were published. CHERI \cite{woodruff_cheri_2014} combines concepts from classic capability systems and RISC architectures, with a strong emphasis on memory protection. RISC-V \cite{asanovic_instruction_2014} is a RISC-based hardware architecture, aimed at providing an extensible open source instruction set architecture (ISA) used as an industry standard by a broad array of hardware vendors. Neither CHERI nor RISC-V were designed with speculative execution vulnerabilities in mind, but Watson \textit{et al.} \cite{watson_capability_2018} observed that CHERI mitigates some aspects of Spectre and Meltdown but is vulnerable to speculative memory access, while Asanovi{\'c} and O'Connor \cite{asanovic_building_2018} announced that RISC-V is not vulnerable because it does not perform speculative memory access. In August 2018, Google announced that the open source implementation of its Titan project, providing a hardware root of trust, will likely be based on RISC-V \cite{rizzo_titan:_2018}. MIT's Sanctum processor \cite{costan_sanctum:_2016} was also based on RISC-V and demonstrated potential for secure hardware partitioning, by adding a small secure CPU to the side of the main CPU. Hardware partitioning might provide a way to mitigate the speculative execution vulnerabilities in multitenant environments, while avoiding major changes to the kernel and operating system. However, genuinely delivering the level of physical isolation that x86 promised would likely require logical partitioning of the main CPU, RAM, and cache of the machine, so the guests and the host operating system could share resources at the hardware level, but be far more restricted at the software level than is currently possible.

The problem of providing secure isolation for containers and virtual machines extends beyond simple refinements to their implementations. When the fundamental assumptions of a system are proven false, then any theorems built on those assumptions may also be false. The secure isolation features of the full stack---from the kernel and operating system, through to virtual machines, containers, and application workloads---are all built on false assumptions about the behavior of the hardware, and will need to be re-examined.

\section{Related implementations} \label{sec-related}

Implementation approaches that adopt the label ``cloud'' \cite{singh_regional_2014, huang_state_2015, lombardi_secure_2011, dildar_effective_2017} are typically virtual machines with added orchestration features to enhance portability. Cloud implementations also tend to favor lighter-weight guest images, which enhances performance and reduces complexity, though cloud images are generally not quite as minimal as container images.

Implementation approaches that adopt the label ``unikernel'' \cite{madhavapeddy_unikernels:_2013, lankes_hermitcore:_2016, williams_unikernel_2016} take minimalist guest images to an extreme, by replacing the kernel and operating system of the guest with a set of highly-optimized libraries that provide the same functionality. The code for an application workload is compiled together with the small subset of unikernel libraries required by the application, resulting in a very small binary that runs directly as a guest image. Historically, unikernels have sacrificed portability of guest images, by targeting only a limited set of virtual machine implementations as their host, but recent work has begun exploring running unikernels as containers \cite{williams_unikernels_2018}. The unikernel approach also reduces the portability of application code, since unikernel frameworks tend to require the application code to be written in the same language as the unikernel libraries.

Implementation approaches that adopt the label ``serverless'' \cite{kanso_serverless:_2017, adzic_serverless_2017, koller_will_2017, hellerstein_serverless_2018} tend to emphasize portability and minimizing complexity. They rely on the underlying infrastructure---typically some combination of bare metal, virtual machines, and/or containers---for whatever secure isolation and performance they provide.

\section{Conclusion} \label{sec-conclusion}

A detailed examination of the history of virtual machines and containers reveals that the two have evolved in tandem from the very beginning. It also reveals that both families of technology are facing significant challenges in providing secure isolation for modern multitenant infrastructures. In light of recent vulnerabilities, patching up existing tools is a necessary and valuable activity in the short-term, but is not sufficient for the long-term. In the coming decades, the computing industry as a whole will need to embrace more radical alternatives in both hardware and software. A deeper understanding of how virtual machines and containers evolved---and the trade-offs made along the way---can lead to new paths of exploration, and help the researchers and developers of today make more informed choices for tomorrow.

\section*{Acknowledgments} \label{sec-acknowledgments}

Thanks to Ravi Nair for help locating and scanning copies of several pivotal IBM papers on virtual machines from the 1960s that were no longer (or perhaps never) available in libraries or online. Thanks also to the reviewers (alphabetically): Matthew Allen, Clint Adams, Ross Anderson, Alastair Beresford, Damian Conway, Kees Cook, Jon Crowcroft, Mike Dodson, Tony Finch, Greg Kroah-Hartman, Anil Madhavapeddy, Ronald Minnich, Richard Mortier, Davanum Srinivas, Tom Sutcliffe, Zahra Tarkhani, and Daniel Thomas. Their feedback was greatly appreciated.

\bibliographystyle{abbrvnat}
\bibliography{article}

\begin{thebibliography}{217}
\providecommand{\natexlab}[1]{#1}
\providecommand{\url}[1]{\texttt{#1}}
\expandafter\ifx\csname urlstyle\endcsname\relax
  \providecommand{\doi}[1]{doi: #1}\else
  \providecommand{\doi}{doi: \begingroup \urlstyle{rm}\Url}\fi

\bibitem[noa(1971)]{noauthor_control_1971}
\emph{Control {Program}-67 {Cambridge} {Monitor} {System}}.
\newblock {IBM} {Type} {III} {Release} {No}. 360D-05.2.005. IBM Corporation,
  Hawthorne, New York, Oct. 1971.

\bibitem[noa(1981)]{noauthor_iapx_1981}
\emph{{iAPX} 432 {General} {Data} {Processor} {Architecture} {Reference}
  {Manual}}.
\newblock Intel Corporation, Aloha, Oregon, 1981.

\bibitem[noa(1997)]{noauthor_protection_1997}
Protection, {Audit} and {Control} {Interfaces}.
\newblock Draft {POSIX} {Standard} 1003.1e, IEEE, Oct. 1997.

\bibitem[noa(2018{\natexlab{a}})]{noauthor_capabilities7_2018}
capabilities(7) man page, Feb. 2018{\natexlab{a}}.
\newblock URL \url{http://man7.org/linux/man-pages/man7/capabilities.7.html}.

\bibitem[noa(2018{\natexlab{b}})]{noauthor_kata_2018}
Kata {Containers} - {The} speed of containers, the security of {VMs}, May
  2018{\natexlab{b}}.
\newblock URL \url{https://katacontainers.io/}.

\bibitem[noa(2018{\natexlab{c}})]{noauthor_linux_2018}
Linux {Containers} - {LXC} - {Introduction}, 2018{\natexlab{c}}.
\newblock URL \url{https://linuxcontainers.org/lxc/introduction/}.

\bibitem[noa(2018{\natexlab{d}})]{noauthor_nabla_2018}
Nabla containers: a new approach to container isolation, Aug.
  2018{\natexlab{d}}.
\newblock URL \url{https://nabla-containers.github.io/}.

\bibitem[noa(2018{\natexlab{e}})]{noauthor_nemu_2018}
{NEMU} - {Modern} {Hypervisor} for the {Cloud}., Dec. 2018{\natexlab{e}}.
\newblock URL \url{https://github.com/intel/nemu}.

\bibitem[noa(2018{\natexlab{f}})]{noauthor_retpoline:_2018}
Retpoline: {A} {Branch} {Target} {Injection} {Mitigation}.
\newblock White {Paper} 337131-003, Intel Corporation, June 2018{\natexlab{f}}.

\bibitem[noa(2018{\natexlab{g}})]{noauthor_speculative_2018}
Speculative {Execution} {Side} {Channel} {Mitigations}.
\newblock Technical Report 336996-003, Intel Corporation, July
  2018{\natexlab{g}}.

\bibitem[noa(2018{\natexlab{h}})]{noauthor_user_namespaces7_2018}
user\_namespaces(7) man page, Feb. 2018{\natexlab{h}}.
\newblock URL
  \url{http://man7.org/linux/man-pages/man7/user_namespaces.7.html}.

\bibitem[noa(2019{\natexlab{a}})]{noauthor_kata_2019}
Kata {Containers} {Architecture}, Jan. 2019{\natexlab{a}}.
\newblock URL \url{https://github.com/kata-containers/documentation}.

\bibitem[noa(2019{\natexlab{b}})]{noauthor_l1tf_2019}
L1tf - {L}1 {Terminal} {Fault} {\textemdash} {The} {Linux} {Kernel}
  documentation, 2019{\natexlab{b}}.
\newblock URL
  \url{https://www.kernel.org/doc/html/latest/admin-guide/l1tf.html}.

\bibitem[Ackerman and Plummer(1967)]{ackerman_implementation_1967}
W.~B. Ackerman and W.~W. Plummer.
\newblock An {Implementation} of a {Multiprocessing} {Computer} {System}.
\newblock In \emph{Proceedings of the {First} {ACM} {Symposium} on {Operating}
  {System} {Principles}}, {SOSP} '67, pages 5.1--5.10, New York, NY, USA, 1967.
  ACM.

\bibitem[Adair et~al.(1966)Adair, Bayles, Comeau, and
  Creasy]{adair_virtual_1966}
R.~J. Adair, R.~U. Bayles, L.~W. Comeau, and R.~J. Creasy.
\newblock A {Virtual} {Machine} {System} for the 360/40.
\newblock Technical Report 36.010, IBM Cambridge Scientific Center, Cambridge,
  MA, USA, May 1966.

\bibitem[Adams and Agesen(2006)]{adams_comparison_2006}
K.~Adams and O.~Agesen.
\newblock A {Comparison} of {Software} and {Hardware} {Techniques} for x86
  {Virtualization}.
\newblock In \emph{Proceedings of the 12th {International} {Conference} on
  {Architectural} {Support} for {Programming} {Languages} and {Operating}
  {Systems}}, {ASPLOS} {XII}, pages 2--13, New York, NY, USA, 2006. ACM.

\bibitem[Adzic and Chatley(2017)]{adzic_serverless_2017}
G.~Adzic and R.~Chatley.
\newblock Serverless {Computing}: {Economic} and {Architectural} {Impact}.
\newblock In \emph{Proceedings of the 2017 11th {Joint} {Meeting} on
  {Foundations} of {Software} {Engineering}}, {ESEC}/{FSE} 2017, pages
  884--889, New York, NY, USA, 2017. ACM.

\bibitem[Ahmad et~al.(2003)Ahmad, Anderson, Holler, Kambo, and
  Makhija]{ahmad_analysis_2003}
I.~Ahmad, J.~M. Anderson, A.~M. Holler, R.~Kambo, and V.~Makhija.
\newblock An analysis of disk performance in {VMware} {ESX} server virtual
  machines.
\newblock In \emph{2003 {IEEE} {International} {Conference} on {Communications}
  ({Cat}. {No}.03CH37441)}, pages 65--76, Oct. 2003.

\bibitem[Amazon(2019)]{amazon_firecracker_2019}
Amazon.
\newblock Firecracker, 2019.
\newblock URL \url{https://firecracker-microvm.github.io/}.

\bibitem[Amdahl et~al.(1964)Amdahl, Blaauw, and
  Brooks]{amdahl_architecture_1964}
G.~M. Amdahl, G.~A. Blaauw, and F.~P. Brooks.
\newblock Architecture of the {IBM} {System}/360.
\newblock \emph{IBM Journal of Research and Development}, 8\penalty0
  (2):\penalty0 87--101, Apr. 1964.

\bibitem[Anderson et~al.(2017)Anderson, Godfrey, and
  Watson]{anderson_towards_2017}
J.~Anderson, S.~Godfrey, and R.~N. Watson.
\newblock Towards oblivious sandboxing with {Capsicum}.
\newblock 2017.

\bibitem[Arcangeli(2005)]{arcangeli_[patch]_2005}
A.~Arcangeli.
\newblock [{PATCH}] seccomp: secure computing support, Mar. 2005.
\newblock URL
  \url{https://git.kernel.org/pub/scm/linux/kernel/git/tglx/history.git/commit/?id=d949d0ec9c601f2b148bed3cdb5f87c052968554}.

\bibitem[Asanovi{\'c} and
  O{\textquoteright}Connor(2018)]{asanovic_building_2018}
K.~Asanovi{\'c} and R.~O{\textquoteright}Connor.
\newblock Building a {More} {Secure} {World} with the {RISC}-{V} {ISA}, Jan.
  2018.
\newblock URL \url{https://riscv.org/2018/01/more-secure-world-risc-v-isa/}.

\bibitem[Asanovi{\'c} and Patterson(2014)]{asanovic_instruction_2014}
K.~Asanovi{\'c} and D.~A. Patterson.
\newblock Instruction {Sets} {Should} {Be} {Free}: {The} {Case} {For}
  {RISC}-{V}.
\newblock Technical Report UCB/EECS-2014-146, EECS Department, University of
  California, Berkeley, Aug. 2014.

\bibitem[Banga et~al.(1999)Banga, Druschel, and Mogul]{banga_resource_1999}
G.~Banga, P.~Druschel, and J.~C. Mogul.
\newblock Resource {Containers}: {A} {New} {Facility} for {Resource}
  {Management} in {Server} {Systems}.
\newblock In \emph{Proceedings of the {Third} {Symposium} on {Operating}
  {Systems} {Design} and {Implementation}}, {OSDI} '99, pages 45--58, Berkeley,
  CA, USA, 1999. USENIX Association.

\bibitem[Barham et~al.(2003)Barham, Dragovic, Fraser, Hand, Harris, Ho,
  Neugebauer, Pratt, and Warfield]{barham_xen_2003}
P.~Barham, B.~Dragovic, K.~Fraser, S.~Hand, T.~Harris, A.~Ho, R.~Neugebauer,
  I.~Pratt, and A.~Warfield.
\newblock Xen and the {Art} of {Virtualization}.
\newblock In \emph{Proceedings of the {Nineteenth} {ACM} {Symposium} on
  {Operating} {Systems} {Principles}}, {SOSP} '03, pages 164--177, New York,
  NY, USA, 2003. ACM.

\bibitem[Barham(1997)]{barham_fresh_1997}
P.~R. Barham.
\newblock A fresh approach to file system quality of service.
\newblock In \emph{Proceedings of 7th {International} {Workshop} on {Network}
  and {Operating} {System} {Support} for {Digital} {Audio} and {Video}
  ({NOSSDAV} '97)}, pages 113--122, May 1997.

\bibitem[Barr(2006)]{barr_amazon_2006}
J.~Barr.
\newblock Amazon {EC}2 {Beta}, Aug. 2006.
\newblock URL \url{https://aws.amazon.com/blogs/aws/amazon_ec2_beta/}.

\bibitem[Bellard(2005)]{bellard_qemu_2005}
F.~Bellard.
\newblock {QEMU}, a {Fast} and {Portable} {Dynamic} {Translator}.
\newblock In \emph{Proceedings of the {USENIX} {Annual} {Technical}
  {Conference}}, {ATEC} '05, pages 41--41, Berkeley, CA, USA, Apr. 2005. USENIX
  Association.

\bibitem[Berger et~al.(2006)Berger, Caceres, Goldman, Perez, Sailer, and van
  Doorn]{berger_vtpm:_2006}
S.~Berger, R.~Caceres, K.~A. Goldman, R.~Perez, R.~Sailer, and L.~van Doorn.
\newblock {vTPM}: {Virtualizing} the {Trusted} {Platform} {Module}.
\newblock In \emph{Proceedings of the 15th {USENIX} {Security} {Symposium}},
  pages 305--320, Vancouver, Canada, Aug. 2006. USENIX Association.

\bibitem[Bernstein(2014)]{bernstein_containers_2014}
D.~Bernstein.
\newblock Containers and {Cloud}: {From} {LXC} to {Docker} to {Kubernetes}.
\newblock \emph{IEEE Cloud Computing}, 1\penalty0 (3):\penalty0 81--84, Sept.
  2014.

\bibitem[Berstis(1980)]{berstis_security_1980}
V.~Berstis.
\newblock Security and {Protection} of {Data} in the {IBM} {System}/38.
\newblock In \emph{Proceedings of the 7th {Annual} {Symposium} on {Computer}
  {Architecture}}, {ISCA} '80, pages 245--252, New York, NY, USA, 1980. ACM.

\bibitem[Biederman(2006)]{biederman_multiple_2006}
E.~W. Biederman.
\newblock Multiple instances of the global linux namespaces.
\newblock In \emph{Proceedings of the {Linux} {Symposium}}, volume~1, pages
  101--112, Ottowa, Canada, 2006.

\bibitem[Bila et~al.(2017)Bila, Dettori, Kanso, Watanabe, and
  Youssef]{bila_leveraging_2017}
N.~Bila, P.~Dettori, A.~Kanso, Y.~Watanabe, and A.~Youssef.
\newblock Leveraging the {Serverless} {Architecture} for {Securing} {Linux}
  {Containers}.
\newblock In \emph{2017 {IEEE} 37th {International} {Conference} on
  {Distributed} {Computing} {Systems} {Workshops} ({ICDCSW})}, pages 401--404,
  June 2017.

\bibitem[Bolte et~al.(2010)Bolte, Sievers, Birkenheuer, Nieh{\"o}rster, and
  Brinkmann]{bolte_non-intrusive_2010}
M.~Bolte, M.~Sievers, G.~Birkenheuer, O.~Nieh{\"o}rster, and A.~Brinkmann.
\newblock Non-intrusive virtualization management using libvirt.
\newblock pages 574--579. European Design and Automation Association, Mar.
  2010.

\bibitem[Boulle(2016)]{boulle_celebrating_2016}
J.~Boulle.
\newblock Celebrating the {Open} {Container} {Initiative} {Image}
  {Specification}, Apr. 2016.
\newblock URL \url{https://coreos.com/blog/oci-image-specification.html}.

\bibitem[Bratus et~al.(2010)Bratus, Locasto, Ramaswamy, and
  Smith]{bratus_vm-based_2010}
S.~Bratus, M.~E. Locasto, A.~Ramaswamy, and S.~W. Smith.
\newblock {VM}-based {Security} {Overkill}: {A} {Lament} for {Applied}
  {Systems} {Security} {Research}.
\newblock In \emph{Proceedings of the 2010 {New} {Security} {Paradigms}
  {Workshop}}, {NSPW} '10, pages 51--60, New York, NY, USA, 2010. ACM.

\bibitem[Brewer(2015)]{brewer_kubernetes_2015}
E.~A. Brewer.
\newblock Kubernetes and the {Path} to {Cloud} {Native}.
\newblock In \emph{Proceedings of the {Sixth} {ACM} {Symposium} on {Cloud}
  {Computing}}, {SoCC} '15, pages 167--167, New York, NY, USA, 2015. ACM.

\bibitem[Bugnion et~al.(1997)Bugnion, Devine, Govil, and
  Rosenblum]{bugnion_disco:_1997}
E.~Bugnion, S.~Devine, K.~Govil, and M.~Rosenblum.
\newblock Disco: {Running} {Commodity} {Operating} {Systems} on {Scalable}
  {Multiprocessors}.
\newblock \emph{ACM Trans. Comput. Syst.}, 15\penalty0 (4):\penalty0 412--447,
  Nov. 1997.

\bibitem[Bugnion et~al.(2012)Bugnion, Devine, Rosenblum, Sugerman, and
  Wang]{bugnion_bringing_2012}
E.~Bugnion, S.~Devine, M.~Rosenblum, J.~Sugerman, and E.~Y. Wang.
\newblock Bringing {Virtualization} to the x86 {Architecture} with the
  {Original} {VMware} {Workstation}.
\newblock \emph{ACM Trans. Comput. Syst.}, 30\penalty0 (4):\penalty0
  12:1--12:51, Nov. 2012.

\bibitem[Bugnion et~al.(2017)Bugnion, Nieh, and Tsafrir]{bugnion_hardware_2017}
E.~Bugnion, J.~Nieh, and D.~Tsafrir.
\newblock \emph{Hardware and {Software} {Support} for {Virtualization}}.
\newblock Morgan \& Claypool, 2017.

\bibitem[Burns et~al.(2016)Burns, Grant, Oppenheimer, Brewer, and
  Wilkes]{burns_borg_2016}
B.~Burns, B.~Grant, D.~Oppenheimer, E.~Brewer, and J.~Wilkes.
\newblock Borg, {Omega}, and {Kubernetes}.
\newblock \emph{Queue}, 14\penalty0 (1):\penalty0 10:70--10:93, Jan. 2016.

\bibitem[Buzen and Gagliardi(1973)]{buzen_evolution_1973}
J.~P. Buzen and U.~O. Gagliardi.
\newblock The {Evolution} of {Virtual} {Machine} {Architecture}.
\newblock In \emph{Proceedings of the {June} 4-8, 1973, {National} {Computer}
  {Conference} and {Exposition}}, {AFIPS} '73, pages 291--299, New York, NY,
  USA, 1973. ACM.

\bibitem[Canella et~al.(2018)Canella, Van~Bulck, Schwarz, Lipp, von Berg,
  Ortner, Piessens, Evtyushkin, and Gruss]{canella_systematic_2018}
C.~Canella, J.~Van~Bulck, M.~Schwarz, M.~Lipp, B.~von Berg, P.~Ortner,
  F.~Piessens, D.~Evtyushkin, and D.~Gruss.
\newblock A {Systematic} {Evaluation} of {Transient} {Execution} {Attacks} and
  {Defenses}.
\newblock \emph{arXiv:1811.05441 [cs]}, Nov. 2018.

\bibitem[Catuogno and Galdi(2016)]{catuogno_evaluation_2016}
L.~Catuogno and C.~Galdi.
\newblock On the {Evaluation} of {Security} {Properties} of {Containerized}
  {Systems}.
\newblock In \emph{2016 15th {International} {Conference} on {Ubiquitous}
  {Computing} and {Communications} and 2016 {International} {Symposium} on
  {Cyberspace} and {Security} ({IUCC}-{CSS})}, pages 69--76, Dec. 2016.

\bibitem[Chandramouli et~al.(2018)Chandramouli, Singhal, Wijesekera, and
  Liu]{chandramouli_methodology_2018}
R.~Chandramouli, A.~Singhal, D.~Wijesekera, and C.~Liu.
\newblock A {Methodology} for {Determining} {Forensic} {Data} {Requirements}
  for {Detecting} {Hypervisor} {Attacks}.
\newblock Technical Report NISTIR 8221 (Draft), National Institute of Standards
  and Technology, Sept. 2018.
\newblock URL
  \url{https://csrc.nist.gov/publications/detail/nistir/8221/draft}.

\bibitem[Chisnall(2007)]{chisnall_definitive_2007}
D.~Chisnall.
\newblock \emph{The {Definitive} {Guide} to the {Xen} {Hypervisor}}.
\newblock Prentice Hall Press, Upper Saddle River, NJ, USA, first edition,
  2007.

\bibitem[Claassen et~al.(2016)Claassen, Koning, and
  Grosso]{claassen_linux_2016}
J.~Claassen, R.~Koning, and P.~Grosso.
\newblock Linux containers networking: {Performance} and scalability of kernel
  modules.
\newblock In \emph{{NOMS} 2016 - 2016 {IEEE}/{IFIP} {Network} {Operations} and
  {Management} {Symposium}}, pages 713--717, Apr. 2016.

\bibitem[Codd(1960{\natexlab{a}})]{codd_multiprogram_1960}
E.~F. Codd.
\newblock Multiprogram {Scheduling}: {Parts} 1 and 2. {Introduction} and
  {Theory}.
\newblock \emph{Communications of the ACM}, 3\penalty0 (6):\penalty0 347--350,
  June 1960{\natexlab{a}}.

\bibitem[Codd(1960{\natexlab{b}})]{codd_multiprogram_1960-1}
E.~F. Codd.
\newblock Multiprogram {Scheduling}: {Parts} 3 and 4. {Scheduling} {Algorithm}
  and {External} {Constraints}.
\newblock \emph{Communications of the ACM}, 3\penalty0 (7):\penalty0 413--418,
  July 1960{\natexlab{b}}.

\bibitem[Codd(1962)]{codd_multiprogramming_1962}
E.~F. Codd.
\newblock Multiprogramming.
\newblock In F.~L. Alt and M.~Rubinoff, editors, \emph{Advances in
  {Computers}}, volume~3, pages 77--153. Elsevier, Jan. 1962.

\bibitem[Codd et~al.(1959)Codd, Lowry, McDonough, and
  Scalzi]{codd_multiprogramming_1959}
E.~F. Codd, E.~S. Lowry, E.~McDonough, and C.~A. Scalzi.
\newblock Multiprogramming {STRETCH}: {Feasibility} {Considerations}.
\newblock \emph{Communications of the ACM}, 2\penalty0 (11):\penalty0 13--17,
  Nov. 1959.

\bibitem[Colp et~al.(2011)Colp, Nanavati, Zhu, Aiello, Coker, Deegan, Loscocco,
  and Warfield]{colp_breaking_2011}
P.~Colp, M.~Nanavati, J.~Zhu, W.~Aiello, G.~Coker, T.~Deegan, P.~Loscocco, and
  A.~Warfield.
\newblock Breaking {Up} is {Hard} to {Do}: {Security} and {Functionality} in a
  {Commodity} {Hypervisor}.
\newblock In \emph{Proceedings of the {Twenty}-{Third} {ACM} {Symposium} on
  {Operating} {Systems} {Principles}}, {SOSP} '11, pages 189--202, New York,
  NY, USA, 2011. ACM.

\bibitem[Combe et~al.(2016)Combe, Martin, and Pietro]{combe_docker_2016}
T.~Combe, A.~Martin, and R.~D. Pietro.
\newblock To {Docker} or {Not} to {Docker}: {A} {Security} {Perspective}.
\newblock \emph{IEEE Cloud Computing}, 3\penalty0 (5):\penalty0 54--62, Sept.
  2016.

\bibitem[Corbat{\'o} et~al.(1962)Corbat{\'o}, Merwin-Daggett, and
  Daley]{corbato_experimental_1962}
F.~J. Corbat{\'o}, M.~Merwin-Daggett, and R.~C. Daley.
\newblock An {Experimental} {Time}-sharing {System}.
\newblock In \emph{Proceedings of the {May} 1-3, 1962, {Spring} {Joint}
  {Computer} {Conference}}, {AIEE}-{IRE} '62 ({Spring}), pages 335--344, New
  York, NY, USA, 1962. ACM.

\bibitem[Corbat{\'o} et~al.(1972)Corbat{\'o}, Saltzer, and
  Clingen]{corbato_multics:_1972}
F.~J. Corbat{\'o}, J.~H. Saltzer, and C.~T. Clingen.
\newblock Multics: {The} {First} {Seven} {Years}.
\newblock In \emph{Proceedings of the {May} 16-18, 1972, {Spring} {Joint}
  {Computer} {Conference}}, {AFIPS} '72 ({Spring}), pages 571--583, New York,
  NY, USA, 1972. ACM.

\bibitem[Corbet(2007)]{corbet_process_2007}
J.~Corbet.
\newblock Process containers, May 2007.
\newblock URL \url{https://lwn.net/Articles/236038/}.

\bibitem[Costan et~al.(2016)Costan, Lebedev, and Devadas]{costan_sanctum:_2016}
V.~Costan, I.~Lebedev, and S.~Devadas.
\newblock Sanctum: {Minimal} {Hardware} {Extensions} for {Strong} {Software}
  {Isolation}.
\newblock pages 857--874, 2016.

\bibitem[Cowan et~al.(2000)Cowan, Beattie, Kroah-Hartman, Pu, Wagle, and
  Gligor]{cowan_subdomain:_2000}
C.~Cowan, S.~Beattie, G.~Kroah-Hartman, C.~Pu, P.~Wagle, and V.~Gligor.
\newblock {SubDomain}: {Parsimonious} {Server} {Security}.
\newblock In \emph{Proceedings of the 14th {USENIX} {Conference} on {System}
  {Administration}}, {LISA} '00, pages 355--368, New Orleans, Louisiana, 2000.
  USENIX Association.

\bibitem[Creasy(1981)]{creasy_origin_1981}
R.~J. Creasy.
\newblock The {Origin} of the {VM}/370 {Time}-{Sharing} {System}.
\newblock \emph{IBM Journal of Research and Development}, 25\penalty0
  (5):\penalty0 483--490, Sept. 1981.

\bibitem[Denning(1976)]{denning_fault_1976}
P.~J. Denning.
\newblock Fault {Tolerant} {Operating} {Systems}.
\newblock \emph{ACM Comput. Surv.}, 8\penalty0 (4):\penalty0 359--389, Dec.
  1976.

\bibitem[Denning(1981)]{denning_performance_1981}
P.~J. Denning.
\newblock Performance {Modeling}: {Experimental} {Computer} {Science} {As}
  {Its} {Best}.
\newblock \emph{Communications of the ACM, President's Letter}, 24\penalty0
  (11):\penalty0 725--727, Nov. 1981.

\bibitem[Dennis(1965)]{dennis_segmentation_1965}
J.~B. Dennis.
\newblock Segmentation and the {Design} of {Multiprogrammed} {Computer}
  {Systems}.
\newblock \emph{Journal of the ACM}, 12\penalty0 (4):\penalty0 589--602, Oct.
  1965.

\bibitem[Dennis and Van~Horn(1966)]{dennis_programming_1966}
J.~B. Dennis and E.~C. Van~Horn.
\newblock Programming {Semantics} for {Multiprogrammed} {Computations}.
\newblock \emph{Communications of the ACM}, 9\penalty0 (3):\penalty0 143--155,
  Mar. 1966.

\bibitem[Dickman(1973)]{dickman_small_1973}
L.~I. Dickman.
\newblock Small {Virtual} {Machines}: {A} {Survey}.
\newblock In \emph{Proceedings of the {Workshop} on {Virtual} {Computer}
  {Systems}}, pages 191--202, New York, NY, USA, 1973. ACM.

\bibitem[Dildar et~al.(2017)Dildar, Khan, Abdullah, and
  Khan]{dildar_effective_2017}
M.~S. Dildar, N.~Khan, J.~B. Abdullah, and A.~S. Khan.
\newblock Effective way to defend the hypervisor attacks in cloud computing.
\newblock In \emph{2017 2nd {International} {Conference} on {Anti}-{Cyber}
  {Crimes} ({ICACC})}, pages 154--159, Mar. 2017.

\bibitem[Drewry(2012)]{drewry_dynamic_2012}
W.~Drewry.
\newblock dynamic seccomp policies (using {BPF} filters), Jan. 2012.
\newblock URL \url{https://lwn.net/Articles/475019/}.

\bibitem[Dunwell(1957)]{dunwell_design_1957}
S.~W. Dunwell.
\newblock Design {Objectives} for the {IBM} {Stretch} {Computer}.
\newblock In \emph{Papers and {Discussions} {Presented} at the {December}
  10-12, 1956, {Eastern} {Joint} {Computer} {Conference}: {New} {Developments}
  in {Computers}}, {AIEE}-{IRE} '56 ({Eastern}), pages 20--22, New York, NY,
  USA, 1957. ACM.

\bibitem[Eckert(1957)]{eckert_univac-larc_1957}
J.~P. Eckert.
\newblock {UNIVAC}-{Larc}, the {Next} {Step} in {Computer} {Design}.
\newblock In \emph{Papers and {Discussions} {Presented} at the {December}
  10-12, 1956, {Eastern} {Joint} {Computer} {Conference}: {New} {Developments}
  in {Computers}}, {AIEE}-{IRE} '56 ({Eastern}), pages 16--20, New York, NY,
  USA, 1957. ACM.

\bibitem[Ei{\ss}feldt(1997)]{eisfeldt_posix:_1997}
H.~Ei{\ss}feldt.
\newblock {POSIX}: {A} {Developer}'s {View} of {Standards}.
\newblock In \emph{Proceedings of the {Annual} {Conference} on {USENIX}
  {Annual} {Technical} {Conference}}, {ATEC} '97, pages 24--24, Berkeley, CA,
  USA, 1997. USENIX Association.

\bibitem[England(1974)]{england_capability_1974}
D.~M. England.
\newblock Capability {Concept} {Mechanism} and {Structure} in {System} 250.
\newblock In \emph{Proceedings of the {International} {Workshop} on
  {Protection} in {Operating} {Systems}}, pages 63--82, France, Aug. 1974.
  Institut de Recherche d{\textquoteright}Informatique et de Automatique
  (IRIA).

\bibitem[Fabry(1967)]{fabry_users_1967}
R.~S. Fabry.
\newblock A user's view of capabilities.
\newblock {ICR} {Quarterly} {Report}~15, University of Chicago, Nov. 1967.

\bibitem[Fabry(1968)]{fabry_preliminary_1968}
R.~S. Fabry.
\newblock Preliminary description of a supervisor for a machine oriented around
  capabilities.
\newblock {ICR} {Quarterly} {Report}~18, University of Chicago, Aug. 1968.

\bibitem[Fabry(1971)]{fabry_list-structured_1971}
R.~S. Fabry.
\newblock \emph{List-structured addressing}.
\newblock {PhD} {Thesis}, University of Chicago, Mar. 1971.

\bibitem[Felter et~al.(2015)Felter, Ferreira, Rajamony, and
  Rubio]{felter_updated_2015}
W.~Felter, A.~Ferreira, R.~Rajamony, and J.~Rubio.
\newblock An updated performance comparison of virtual machines and {Linux}
  containers.
\newblock In \emph{2015 {IEEE} {International} {Symposium} on {Performance}
  {Analysis} of {Systems} and {Software} ({ISPASS})}, pages 171--172, Mar.
  2015.

\bibitem[Frankovich and Peterson(1957)]{frankovich_functional_1957}
J.~M. Frankovich and H.~P. Peterson.
\newblock A {Functional} {Description} of the {Lincoln} {TX}-2 {Computer}.
\newblock In \emph{Papers {Presented} at the {February} 26-28, 1957, {Western}
  {Joint} {Computer} {Conference}: {Techniques} for {Reliability}},
  {IRE}-{AIEE}-{ACM} '57 ({Western}), pages 146--155, New York, NY, USA, 1957.
  ACM.

\bibitem[Galley(1973)]{galley_pdp-10_1973}
S.~W. Galley.
\newblock {PDP}-10 virtual machines.
\newblock pages 30--34. ACM, Mar. 1973.

\bibitem[Gao et~al.(2017)Gao, Gu, Kayaalp, Pendarakis, and
  Wang]{gao_containerleaks:_2017}
X.~Gao, Z.~Gu, M.~Kayaalp, D.~Pendarakis, and H.~Wang.
\newblock {ContainerLeaks}: {Emerging} {Security} {Threats} of {Information}
  {Leakages} in {Container} {Clouds}.
\newblock In \emph{2017 47th {Annual} {IEEE}/{IFIP} {International}
  {Conference} on {Dependable} {Systems} and {Networks} ({DSN})}, pages
  237--248, June 2017.

\bibitem[Garfinkel and Rosenblum(2003)]{garfinkel_virtual_2003}
T.~Garfinkel and M.~Rosenblum.
\newblock A {Virtual} {Machine} {Introspection} {Based} {Architecture} for
  {Intrusion} {Detection}.
\newblock In \emph{Proceedings of the {Network} and {Distributed} {Systems}
  {Security} {Symposium}}, volume~1, pages 253--285, 2003.

\bibitem[Garfinkel et~al.(2007)Garfinkel, Adams, Warfield, and
  Franklin]{garfinkel_compatibility_2007}
T.~Garfinkel, K.~Adams, A.~Warfield, and J.~Franklin.
\newblock Compatibility is {Not} {Transparency}: {VMM} {Detection} {Myths} and
  {Realities}.
\newblock In \emph{Proceedings of the 11th {USENIX} {Workshop} on {Hot}
  {Topics} in {Operating} {Systems}}, {HOTOS}'07, page~6, Berkeley, CA, USA,
  2007. USENIX Association.

\bibitem[Gill(1958)]{gill_parallel_1958}
S.~Gill.
\newblock Parallel {Programming}.
\newblock \emph{The Computer Journal}, 1\penalty0 (1):\penalty0 2--10, Jan.
  1958.

\bibitem[Goldberg(1972)]{goldberg_architectural_1972}
R.~P. Goldberg.
\newblock \emph{Architectural {Principles} for {Virtual} {Computer} {Systems}}.
\newblock {PhD} {Thesis}, Harvard University, Cambridge, MA, 1972.

\bibitem[Goldberg(1973)]{goldberg_architecture_1973}
R.~P. Goldberg.
\newblock Architecture of {Virtual} {Machines}.
\newblock In \emph{Proceedings of the {Workshop} on {Virtual} {Computer}
  {Systems}}, pages 74--112, New York, NY, USA, 1973. ACM.

\bibitem[Goldberg(1974)]{goldberg_survey_1974}
R.~P. Goldberg.
\newblock Survey of {Virtual} {Machine} {Research}.
\newblock \emph{Computer}, 7\penalty0 (6):\penalty0 34--45, June 1974.

\bibitem[Google(2018{\natexlab{a}})]{google_chrome_2018}
Google.
\newblock Chrome {OS} {Virtual} {Machine} {Monitor}, Sept. 2018{\natexlab{a}}.
\newblock URL
  \url{https://chromium.googlesource.com/chromiumos/platform/crosvm/}.

\bibitem[Google(2018{\natexlab{b}})]{google_fuchsia_2018}
Google.
\newblock Fuchsia is not {Linux}: {A} modular, capability-based operating
  system, Oct. 2018{\natexlab{b}}.
\newblock URL
  \url{https://fuchsia.googlesource.com/docs/+/HEAD/the-book/README.md}.

\bibitem[Google(2019)]{google_gvisor_2019}
Google.
\newblock {gVisor} - {Container} {Runtime} {Sandbox}, Feb. 2019.
\newblock URL \url{https://github.com/google/gvisor}.

\bibitem[Graber(2016)]{graber_lxd_2016}
S.~Graber.
\newblock {LXD} 2.0, Mar. 2016.
\newblock URL
  \url{https://stgraber.org/2016/03/11/lxd-2-0-blog-post-series-012/}.

\bibitem[Gr{\"u}nbacher(2003)]{grunbacher_posix_2003}
A.~Gr{\"u}nbacher.
\newblock {POSIX} {Access} {Control} {Lists} on {Linux}.
\newblock In \emph{Proceedings of the 2003 {USENIX} {Annual} {Technical}
  {Conference}}, pages 259--272, San Antonio, Texas, June 2003. USENIX
  Association.

\bibitem[Hansen et~al.(1982)Hansen, Linton, Mayo, Murphy, and
  Patterson]{hansen_performance_1982}
P.~M. Hansen, M.~A. Linton, R.~N. Mayo, M.~Murphy, and D.~A. Patterson.
\newblock A {Performance} {Evaluation} of the {Intel} {iAPX} 432.
\newblock \emph{SIGARCH Comput. Archit. News}, 10\penalty0 (4):\penalty0
  17--26, June 1982.

\bibitem[Hellerstein et~al.(2018)Hellerstein, Faleiro, Gonzalez,
  Schleier-Smith, Sreekanti, Tumanov, and Wu]{hellerstein_serverless_2018}
J.~M. Hellerstein, J.~Faleiro, J.~E. Gonzalez, J.~Schleier-Smith, V.~Sreekanti,
  A.~Tumanov, and C.~Wu.
\newblock Serverless {Computing}: {One} {Step} {Forward}, {Two} {Steps} {Back}.
\newblock \emph{arXiv:1812.03651 [cs]}, Dec. 2018.

\bibitem[Hennessy(2018)]{hennessy_era_2018}
J.~Hennessy.
\newblock The era of security: {Introduction}.
\newblock In \emph{Proceedings of the 2018 {IEEE} {Hot} {Chips} {Symposium}},
  Cupertino, CA, Aug. 2018. IEEE.
\newblock URL \url{https://youtu.be/d5XzVF0sAZo}.

\bibitem[Hosseinzadeh et~al.(2016)Hosseinzadeh, Laur{\'e}n, and
  Lepp{\"a}nen]{hosseinzadeh_security_2016}
S.~Hosseinzadeh, S.~Laur{\'e}n, and V.~Lepp{\"a}nen.
\newblock Security in {Container}-based {Virtualization} {Through} {vTPM}.
\newblock In \emph{Proceedings of the 9th {International} {Conference} on
  {Utility} and {Cloud} {Computing}}, {UCC} '16, pages 214--219, New York, NY,
  USA, 2016. ACM.

\bibitem[Houdek et~al.(1981)Houdek, Soltis, and Hoffman]{houdek_ibm_1981}
M.~E. Houdek, F.~G. Soltis, and R.~L. Hoffman.
\newblock {IBM} {System}/38 {Support} for {Capability}-based {Addressing}.
\newblock In \emph{Proceedings of the 8th {Annual} {Symposium} on {Computer}
  {Architecture}}, {ISCA} '81, pages 341--348, Los Alamitos, CA, USA, 1981.
  IEEE Computer Society Press.

\bibitem[Huang et~al.(2015)Huang, Ganjali, Kim, Oh, and Lie]{huang_state_2015}
W.~Huang, A.~Ganjali, B.~H. Kim, S.~Oh, and D.~Lie.
\newblock The {State} of {Public} {Infrastructure}-as-a-{Service} {Cloud}
  {Security}.
\newblock \emph{ACM Comput. Surv.}, 47\penalty0 (4):\penalty0 68:1--68:31, June
  2015.

\bibitem[Huang et~al.(2008)Huang, Stavrou, Ghosh, and
  Jajodia]{huang_efficiently_2008}
Y.~Huang, A.~Stavrou, A.~K. Ghosh, and S.~Jajodia.
\newblock Efficiently {Tracking} {Application} {Interactions} {Using}
  {Lightweight} {Virtualization}.
\newblock In \emph{Proceedings of the 1st {ACM} {Workshop} on {Virtual}
  {Machine} {Security}}, {VMSec} '08, pages 19--28, New York, NY, USA, 2008.
  ACM.

\bibitem[Hykes(2014)]{hykes_docker_2014}
S.~Hykes.
\newblock Docker 0.9: introducing execution drivers and libcontainer, Apr.
  2014.
\newblock URL
  \url{https://blog.docker.com/2014/03/docker-0-9-introducing-execution-drivers-and-libcontainer/}.

\bibitem[Hykes(2015)]{hykes_introducing_2015}
S.~Hykes.
\newblock Introducing {runC}: a lightweight universal container runtime, June
  2015.
\newblock URL \url{https://blog.docker.com/2015/06/runc/}.

\bibitem[Ishiguro and Kono(2018)]{ishiguro_hardening_2018}
K.~Ishiguro and K.~Kono.
\newblock Hardening {Hypervisors} {Against} {Vulnerabilities} in {Instruction}
  {Emulators}.
\newblock In \emph{Proceedings of the 11th {European} {Workshop} on {Systems}
  {Security}}, {EuroSec}'18, pages 7:1--7:6, New York, NY, USA, 2018. ACM.

\bibitem[Jian and Chen(2017)]{jian_defense_2017}
Z.~Jian and L.~Chen.
\newblock A {Defense} {Method} {Against} {Docker} {Escape} {Attack}.
\newblock In \emph{Proceedings of the 2017 {International} {Conference} on
  {Cryptography}, {Security} and {Privacy}}, {ICCSP} '17, pages 142--146, New
  York, NY, USA, 2017. ACM.

\bibitem[Jones et~al.(1979)Jones, Chansler, Durham, Schwans, and
  Vegdahl]{jones_staros_1979}
A.~K. Jones, R.~J. Chansler, Jr., I.~Durham, K.~Schwans, and S.~R. Vegdahl.
\newblock {StarOS}, a {Multiprocessor} {Operating} {System} for the {Support}
  of {Task} {Forces}.
\newblock In \emph{Proceedings of the {Seventh} {ACM} {Symposium} on
  {Operating} {Systems} {Principles}}, {SOSP} '79, pages 117--127, New York,
  NY, USA, 1979. ACM.

\bibitem[Joy(2015)]{joy_performance_2015}
A.~M. Joy.
\newblock Performance comparison between {Linux} containers and virtual
  machines.
\newblock In \emph{2015 {International} {Conference} on {Advances} in
  {Computer} {Engineering} and {Applications}}, pages 342--346, Mar. 2015.

\bibitem[Kamp and Watson(2000)]{kamp_jails:_2000}
P.-H. Kamp and R.~N.~M. Watson.
\newblock Jails: {Confining} the omnipotent root.
\newblock In \emph{Proceedings of the 2nd {International} {SANE} {Conference}},
  Maastricht, The Netherlands, 2000.

\bibitem[Kanso and Youssef(2017)]{kanso_serverless:_2017}
A.~Kanso and A.~Youssef.
\newblock Serverless: {Beyond} the {Cloud}.
\newblock In \emph{Proceedings of the 2Nd {International} {Workshop} on
  {Serverless} {Computing}}, {WoSC} '17, pages 6--10, New York, NY, USA, 2017.
  ACM.

\bibitem[Kappel et~al.(2009)Kappel, Velte, and Velte]{kappel_microsoft_2009}
J.~A. Kappel, A.~Velte, and T.~Velte.
\newblock \emph{Microsoft {Virtualization} with {Hyper}-{V}: {Manage} {Your}
  {Datacenter} with {Hyper}-{V}, {Virtual} {PC}, {Virtual} {Server}, and
  {Application} {Virtualization}}.
\newblock McGraw Hill Professional, Sept. 2009.

\bibitem[Kernighan and McIlroy(1979)]{kernighan_unix_1979}
B.~Kernighan and M.~McIlroy.
\newblock \emph{{UNIX} {Time}-sharing {System}: {UNIX} {Programmer}'s
  {Manual}}, volume~1.
\newblock Bell Telephone Laboratories, Incorporated, Murray Hill, New Jersey,
  7th edition, 1979.

\bibitem[Kerrisk(2013)]{kerrisk_namespaces_2013}
M.~Kerrisk.
\newblock Namespaces in operation, part 1: namespaces overview, Jan. 2013.
\newblock URL \url{https://lwn.net/Articles/531114/}.

\bibitem[Kivity et~al.(2007)Kivity, Kamay, Laor, Lublin, and
  Liguori]{kivity_kvm:_2007}
A.~Kivity, Y.~Kamay, D.~Laor, U.~Lublin, and A.~Liguori.
\newblock {KVM}: the {Linux} {Virtual} {Machine} {Monitor}.
\newblock In \emph{In {Proceedings} of the 2007 {Ottawa} {Linux} {Symposium}
  ({OLS}{\textquoteright}-07}, 2007.

\bibitem[Kocher et~al.(2018)Kocher, Genkin, Gruss, Haas, Hamburg, Lipp,
  Mangard, Prescher, Schwarz, and Yarom]{kocher_spectre_2018}
P.~Kocher, D.~Genkin, D.~Gruss, W.~Haas, M.~Hamburg, M.~Lipp, S.~Mangard,
  T.~Prescher, M.~Schwarz, and Y.~Yarom.
\newblock Spectre {Attacks}: {Exploiting} {Speculative} {Execution}.
\newblock \emph{arXiv:1801.01203 [cs]}, Jan. 2018.

\bibitem[Kogut(1973)]{kogut_segment_1973}
R.~M. Kogut.
\newblock The {Segment} {Based} {File} {Support} {System}.
\newblock pages 35--42. ACM, Mar. 1973.

\bibitem[Koller and Williams(2017)]{koller_will_2017}
R.~Koller and D.~Williams.
\newblock Will {Serverless} {End} the {Dominance} of {Linux} in the {Cloud}?
\newblock In \emph{Proceedings of the 16th {Workshop} on {Hot} {Topics} in
  {Operating} {Systems}}, {HotOS} '17, pages 169--173, New York, NY, USA, 2017.
  ACM Press.

\bibitem[Kov{\'a}cs(2017)]{kovacs_comparison_2017}
{\'A}.~Kov{\'a}cs.
\newblock Comparison of different {Linux} containers.
\newblock In \emph{2017 40th {International} {Conference} on
  {Telecommunications} and {Signal} {Processing} ({TSP})}, pages 47--51, July
  2017.

\bibitem[Krude and Meyer(2013)]{krude_versatile_2013}
J.~Krude and U.~Meyer.
\newblock A {Versatile} {Code} {Execution} {Isolation} {Framework} with
  {Security} {First}.
\newblock In \emph{Proceedings of the 2013 {ACM} {Workshop} on {Cloud}
  {Computing} {Security} {Workshop}}, {CCSW} '13, pages 1--10, New York, NY,
  USA, 2013. ACM.

\bibitem[Lampson and Sturgis(1976)]{lampson_reflections_1976}
B.~W. Lampson and H.~E. Sturgis.
\newblock Reflections on an {Operating} {System} {Design}.
\newblock \emph{Commun. ACM}, 19\penalty0 (5):\penalty0 251--265, May 1976.

\bibitem[Lankes et~al.(2016)Lankes, Pickartz, and
  Breitbart]{lankes_hermitcore:_2016}
S.~Lankes, S.~Pickartz, and J.~Breitbart.
\newblock {HermitCore}: {A} {Unikernel} for {Extreme} {Scale} {Computing}.
\newblock In \emph{Proceedings of the 6th {International} {Workshop} on
  {Runtime} and {Operating} {Systems} for {Supercomputers}}, {ROSS} '16, pages
  4:1--4:8, New York, NY, USA, 2016. ACM.

\bibitem[Lauer and Snow(1972)]{lauer_is_1972}
H.~C. Lauer and C.~R. Snow.
\newblock Is {Supervisor}-{State} {Necessary}?
\newblock In \emph{Procedings of the {ACM} {AICA} {International} {Computing}
  {Symposium}}, Venice, Italy, 1972. University of Newcastle upon Tyne,
  Computing Laboratory.

\bibitem[Lauer and Wyeth(1973)]{lauer_recursive_1973}
H.~C. Lauer and D.~Wyeth.
\newblock A recursive virtual machine architecture.
\newblock pages 113--116. ACM, Mar. 1973.

\bibitem[Levy(1984)]{levy_capability-based_1984}
H.~M. Levy.
\newblock \emph{Capability-{Based} {Computer} {Systems}}.
\newblock Digital Press, Newton, MA, USA, 1984.

\bibitem[Li et~al.(2017)Li, Kihl, Lu, and Andersson]{li_performance_2017}
Z.~Li, M.~Kihl, Q.~Lu, and J.~A. Andersson.
\newblock Performance {Overhead} {Comparison} between {Hypervisor} and
  {Container} {Based} {Virtualization}.
\newblock In \emph{2017 {IEEE} 31st {International} {Conference} on {Advanced}
  {Information} {Networking} and {Applications} ({AINA})}, pages 955--962, Mar.
  2017.

\bibitem[Liguori(2012)]{liguori_qemu_2012}
A.~Liguori.
\newblock {QEMU} 1.3.0 release, Dec. 2012.
\newblock URL
  \url{https://lists.gnu.org/archive/html/qemu-devel/2012-12/msg00123.html}.

\bibitem[Lipner et~al.(1974)Lipner, Wulf, Schell, Popek, Neumann, Weissman, and
  Linden]{lipner_security_1974}
S.~B. Lipner, W.~A. Wulf, R.~R. Schell, G.~J. Popek, P.~G. Neumann,
  C.~Weissman, and T.~A. Linden.
\newblock Security {Kernels}.
\newblock In \emph{Proceedings of the {AFIPS} {National} {Computer}
  {Conference}}, {AFIPS} '74, pages 973--980, New York, NY, USA, 1974. ACM.

\bibitem[Lipp et~al.(2018)Lipp, Schwarz, Gruss, Prescher, Haas, Mangard,
  Kocher, Genkin, Yarom, and Hamburg]{lipp_meltdown_2018}
M.~Lipp, M.~Schwarz, D.~Gruss, T.~Prescher, W.~Haas, S.~Mangard, P.~Kocher,
  D.~Genkin, Y.~Yarom, and M.~Hamburg.
\newblock Meltdown.
\newblock \emph{arXiv:1801.01207 [cs]}, Jan. 2018.

\bibitem[Lombardi and Di~Pietro(2011)]{lombardi_secure_2011}
F.~Lombardi and R.~Di~Pietro.
\newblock Secure virtualization for cloud computing.
\newblock \emph{Journal of Network and Computer Applications}, 34\penalty0
  (4):\penalty0 1113--1122, July 2011.

\bibitem[Loscocco and Smalley(2001)]{loscocco_integrating_2001}
P.~Loscocco and S.~Smalley.
\newblock Integrating {Flexible} {Support} for {Security} {Policies} into the
  {Linux} {Operating} {System}.
\newblock In \emph{Proceedings of the {FREENIX} {Track}: 2001 {USENIX} {Annual}
  {Technical} {Conference}}, pages 29--42, Berkeley, CA, USA, June 2001. USENIX
  Association.

\bibitem[Lottiaux and Morin(2001)]{lottiaux_containers:_2001}
R.~Lottiaux and C.~Morin.
\newblock Containers: a sound basis for a true single system image.
\newblock In \emph{Proceedings {First} {IEEE}/{ACM} {International} {Symposium}
  on {Cluster} {Computing} and the {Grid}}, pages 66--73, May 2001.

\bibitem[Luzzardi(2015)]{luzzardi_announcing_2015}
A.~Luzzardi.
\newblock Announcing {Swarm} 1.0: {Production}-ready clustering at any scale,
  Nov. 2015.
\newblock URL \url{https://blog.docker.com/2015/11/swarm-1-0/}.

\bibitem[Madhavapeddy et~al.(2013)Madhavapeddy, Mortier, Rotsos, Scott, Singh,
  Gazagnaire, Smith, Hand, and Crowcroft]{madhavapeddy_unikernels:_2013}
A.~Madhavapeddy, R.~Mortier, C.~Rotsos, D.~Scott, B.~Singh, T.~Gazagnaire,
  S.~Smith, S.~Hand, and J.~Crowcroft.
\newblock Unikernels: library operating systems for the cloud.
\newblock In \emph{Proceedings of the {Eighteenth} {International} {Conference}
  on {Architectural} {Support} for {Programming} {Languages} and {Operating}
  {Systems}}, {ASPLOS} '13, pages 461--472, New York, NY, USA, 2013. ACM.

\bibitem[Madnick and Donovan(1973)]{madnick_application_1973}
S.~E. Madnick and J.~J. Donovan.
\newblock Application and {Analysis} of the {Virtual} {Machine} {Approach} to
  {Information} {System} {Security} and {Isolation}.
\newblock In \emph{Proceedings of the {Workshop} on {Virtual} {Computer}
  {Systems}}, pages 210--224, New York, NY, USA, 1973. ACM.

\bibitem[Manco et~al.(2017)Manco, Lupu, Schmidt, Mendes, Kuenzer, Sati,
  Yasukata, Raiciu, and Huici]{manco_my_2017}
F.~Manco, C.~Lupu, F.~Schmidt, J.~Mendes, S.~Kuenzer, S.~Sati, K.~Yasukata,
  C.~Raiciu, and F.~Huici.
\newblock My {VM} is {Lighter} (and {Safer}) {Than} {Your} {Container}.
\newblock In \emph{Proceedings of the 26th {Symposium} on {Operating} {Systems}
  {Principles}}, {SOSP} '17, pages 218--233, New York, NY, USA, 2017. ACM.

\bibitem[Margery et~al.(2004)Margery, Lottiaux, and
  Morin]{margery_capabilities_2004}
D.~Margery, R.~Lottiaux, and C.~Morin.
\newblock Capabilities for per {Process} {Tuning} of {Distributed} {Operating}
  {Systems}.
\newblock Research {Report} RR-5411, INRIA, 2004.

\bibitem[Martin et~al.(2018)Martin, Raponi, Combe, and
  Di~Pietro]{martin_docker_2018}
A.~Martin, S.~Raponi, T.~Combe, and R.~Di~Pietro.
\newblock Docker ecosystem {\textendash} {Vulnerability} {Analysis}.
\newblock \emph{Computer Communications}, 122:\penalty0 30--43, June 2018.

\bibitem[Mattetti et~al.(2015)Mattetti, Shulman-Peleg, Allouche, Corradi,
  Dolev, and Foschini]{mattetti_securing_2015}
M.~Mattetti, A.~Shulman-Peleg, Y.~Allouche, A.~Corradi, S.~Dolev, and
  L.~Foschini.
\newblock Securing the infrastructure and the workloads of linux containers.
\newblock In \emph{2015 {IEEE} {Conference} on {Communications} and {Network}
  {Security} ({CNS})}, pages 559--567, Sept. 2015.

\bibitem[Matthews et~al.(2007)Matthews, Hu, Hapuarachchi, Deshane, Dimatos,
  Hamilton, McCabe, and Owens]{matthews_quantifying_2007}
J.~N. Matthews, W.~Hu, M.~Hapuarachchi, T.~Deshane, D.~Dimatos, G.~Hamilton,
  M.~McCabe, and J.~Owens.
\newblock Quantifying the {Performance} {Isolation} {Properties} of
  {Virtualization} {Systems}.
\newblock In \emph{Proceedings of the 2007 {Workshop} on {Experimental}
  {Computer} {Science}}, {ExpCS} '07, New York, NY, USA, 2007. ACM.

\bibitem[Mayer(1982)]{mayer_architecture_1982}
A.~J.~W. Mayer.
\newblock The {Architecture} of the {Burroughs} {B}5000: 20 {Years} {Later} and
  {Still} {Ahead} of the {Times}?
\newblock \emph{SIGARCH Comput. Archit. News}, 10\penalty0 (4):\penalty0 3--10,
  June 1982.

\bibitem[Mazor(2010)]{mazor_intels_2010}
S.~Mazor.
\newblock Intel's 8086.
\newblock \emph{IEEE Annals of the History of Computing}, 32\penalty0
  (1):\penalty0 75--79, Jan. 2010.

\bibitem[McKusick(1999)]{mckusick_twenty_1999}
M.~K. McKusick.
\newblock Twenty {Years} of {Berkeley} {Unix} - {From} {AT}\&{T}-{Owned} to
  {Freely} {Redistributable}.
\newblock In \emph{Open {Sources}: {Voices} from the {Open} {Source}
  {Revolution}}. O'Reilly Media, Inc., Jan. 1999.

\bibitem[McKusick et~al.(1989)McKusick, Karels, Sklower, Fall, Teitelbaum, and
  Bostic]{mckusick_current_1989}
M.~K. McKusick, M.~J. Karels, K.~Sklower, K.~Fall, M.~Teitelbaum, and
  K.~Bostic.
\newblock Current {Research} by {The} {Computer} {Systems} {Research} {Group}
  of {Berkeley}.
\newblock In \emph{Proceedings of the {European} {UNIX} {Users} {Group}},
  Brussels, Belgium, Apr. 1989.

\bibitem[McKusick et~al.(2014)McKusick, Neville-Neil, and
  Watson]{mckusick_design_2014}
M.~K. McKusick, G.~V. Neville-Neil, and R.~N.~M. Watson.
\newblock \emph{The {Design} and {Implementation} of the {FreeBSD} {Operating}
  {System}}.
\newblock Addison-Wesley Professional, 2nd edition edition, Sept. 2014.

\bibitem[Merkel(2014)]{merkel_docker:_2014}
D.~Merkel.
\newblock Docker: {Lightweight} {Linux} {Containers} for {Consistent}
  {Development} and {Deployment}.
\newblock \emph{Linux Journal}, 2014\penalty0 (239), Mar. 2014.

\bibitem[Miller and Chen(2012)]{miller_exercise_2012}
A.~Miller and L.~Chen.
\newblock An {Exercise} in {Secure} {High} {Performance} {Virtual}
  {Containers}.
\newblock page~5, Las Vegas, NV, USA, July 2012.

\bibitem[Miller et~al.(2003)Miller, Yee, and Shapiro]{miller_capability_2003}
M.~S. Miller, K.-P. Yee, and J.~Shapiro.
\newblock Capability {Myths} {Demolished}.
\newblock Technical {Report} SRL2003-02, Johns Hopkins University, Systems
  Research Laboratory, Baltimore, Maryland, 2003.

\bibitem[Morabito et~al.(2015)Morabito, Kj{\"a}llman, and
  Komu]{morabito_hypervisors_2015}
R.~Morabito, J.~Kj{\"a}llman, and M.~Komu.
\newblock Hypervisors vs. {Lightweight} {Virtualization}: {A} {Performance}
  {Comparison}.
\newblock In \emph{2015 {IEEE} {International} {Conference} on {Cloud}
  {Engineering}}, pages 386--393, Mar. 2015.

\bibitem[Morin et~al.(2002)Morin, Gallard, Lottiaux, and
  Vallee]{morin_towards_2002}
C.~Morin, P.~Gallard, R.~Lottiaux, and G.~Vallee.
\newblock Towards an efficient single system image cluster operating system.
\newblock In \emph{Fifth {International} {Conference} on {Algorithms} and
  {Architectures} for {Parallel} {Processing}, 2002. {Proceedings}.}, pages
  370--377, Oct. 2002.

\bibitem[Nagar et~al.(2003)Nagar, Franke, Choi, Seetharaman, Kaplan, Singhvi,
  Kashyap, and Kravetz]{nagar_class-based_2003}
S.~Nagar, H.~Franke, J.~Choi, C.~Seetharaman, S.~Kaplan, N.~Singhvi,
  V.~Kashyap, and M.~Kravetz.
\newblock Class-based {Prioritized} {Resource} {Control} in {Linux}.
\newblock In \emph{Proceedings of the {Linux} {Symposium}}, page~21, Ottowa,
  Canada, July 2003.

\bibitem[Nanba et~al.(1985)Nanba, Ohno, Kubo, Morisue, Ohshima, and
  Yamagishi]{nanba_vm/4:_1985}
S.~Nanba, N.~Ohno, H.~Kubo, H.~Morisue, T.~Ohshima, and H.~Yamagishi.
\newblock {VM}/4: {ACOS}-4 {Virtual} {Machine} {Architecture}.
\newblock In \emph{Proceedings of the 12th {Annual} {International} {Symposium}
  on {Computer} {Architecture}}, {ISCA} '85, pages 171--178, Los Alamitos, CA,
  USA, 1985. IEEE Computer Society Press.

\bibitem[Needham and Walker(1977)]{needham_cambridge_1977}
R.~M. Needham and R.~D.~H. Walker.
\newblock The {Cambridge} {CAP} {Computer} and its protection system.
\newblock In \emph{Proceedings of the {Sixth} {ACM} {Symposium} on {Operating}
  {Systems} {Principles}}, pages 1--10, New York, NY, USA, Nov. 1977. ACM.

\bibitem[Nelson(1964)]{nelson_mapping_1964}
R.~A. Nelson.
\newblock Mapping {Devices} and the {M}44 {Data} {Processing} {System}.
\newblock Research {Report} RC-1303, IBM Thomas J. Watson Research Center,
  Yorktown Heights, NY, 1964.

\bibitem[Neumann(1980)]{neumann_provably_1980}
P.~G. Neumann.
\newblock A {Provably} {Secure} {Operating} {System}: {The} system, its
  applications, and proofs.
\newblock Technical report, Computer Science Laboratory, SRI International,
  1980.

\bibitem[Neumann and Feiertag(2003)]{neumann_psos_2003}
P.~G. Neumann and R.~J. Feiertag.
\newblock {PSOS} revisited.
\newblock In \emph{Proceedings of the 19th {Annual} {Computer} {Security}
  {Applications} {Conference}}, pages 208--216, Dec. 2003.

\bibitem[Norton(2016)]{norton_hardware_2016}
R.~M. Norton.
\newblock Hardware support for compartmentalisation.
\newblock Technical Report UCAM-CL-TR-887, University of Cambridge, Computer
  Laboratory, May 2016.

\bibitem[Opler and Baird(1959)]{opler_multiprogramming:_1959}
A.~Opler and N.~Baird.
\newblock Multiprogramming: {The} {Programmer}'s {View}.
\newblock In \emph{Preprints of {Papers} {Presented} at the 14th {National}
  {Meeting} of the {Association} for {Computing} {Machinery}}, {ACM} '59, pages
  1--4, New York, NY, USA, 1959. ACM.

\bibitem[Osman et~al.(2002)Osman, Subhraveti, Su, and Nieh]{osman_design_2002}
S.~Osman, D.~Subhraveti, G.~Su, and J.~Nieh.
\newblock The {Design} and {Implementation} of {Zap}: {A} {System} for
  {Migrating} {Computing} {Environments}.
\newblock In \emph{Proceedings of the 5th {Operating} {Systems} {Design} and
  {Implementation} ({OSDI})}, Boston, MA, Dec. 2002.

\bibitem[Parmelee et~al.(1972)Parmelee, Peterson, Tillman, and
  Hatfield]{parmelee_virtual_1972}
R.~P. Parmelee, T.~I. Peterson, C.~C. Tillman, and D.~J. Hatfield.
\newblock Virtual storage and virtual machine concepts.
\newblock \emph{IBM Systems Journal}, 11\penalty0 (2):\penalty0 99--130, 1972.

\bibitem[Patterson and Ditzel(1980)]{patterson_case_1980}
D.~A. Patterson and D.~R. Ditzel.
\newblock The {Case} for the {Reduced} {Instruction} {Set} {Computer}.
\newblock \emph{SIGARCH Comput. Archit. News}, 8\penalty0 (6):\penalty0 25--33,
  Oct. 1980.

\bibitem[Patterson and Sequin(1981)]{patterson_risc_1981}
D.~A. Patterson and C.~H. Sequin.
\newblock {RISC} {I}: {A} {Reduced} {Instruction} {Set} {VLSI} {Computer}.
\newblock In \emph{Proceedings of the 8th {Annual} {Symposium} on {Computer}
  {Architecture}}, {ISCA} '81, pages 443--457, Los Alamitos, CA, USA, 1981.
  IEEE Computer Society Press.

\bibitem[Pearce et~al.(2013)Pearce, Zeadally, and
  Hunt]{pearce_virtualization:_2013}
M.~Pearce, S.~Zeadally, and R.~Hunt.
\newblock Virtualization: {Issues}, security threats, and solutions.
\newblock \emph{ACM Computing Surveys}, 45\penalty0 (2):\penalty0 1--39, Feb.
  2013.

\bibitem[Perez-Botero et~al.(2013)Perez-Botero, Szefer, and
  Lee]{perez-botero_characterizing_2013}
D.~Perez-Botero, J.~Szefer, and R.~B. Lee.
\newblock Characterizing {Hypervisor} {Vulnerabilities} in {Cloud} {Computing}
  {Servers}.
\newblock In \emph{Proceedings of the 2013 {International} {Workshop} on
  {Security} in {Cloud} {Computing}}, Cloud {Computing} '13, pages 3--10, New
  York, NY, USA, 2013. ACM.

\bibitem[Pike et~al.(1993)Pike, Presotto, Thompson, Trickey, and
  Winterbottom]{pike_use_1993}
R.~Pike, D.~Presotto, K.~Thompson, H.~Trickey, and P.~Winterbottom.
\newblock The {Use} of {Name} {Spaces} in {Plan} 9.
\newblock \emph{SIGOPS Oper. Syst. Rev.}, 27\penalty0 (2):\penalty0 72--76,
  Apr. 1993.

\bibitem[Popek and Kline(1975)]{popek_verifiable_1975}
G.~Popek and C.~Kline.
\newblock A verifiable protection system.
\newblock \emph{ACM SIGPLAN Notices}, 10\penalty0 (6):\penalty0 294--304, June
  1975.

\bibitem[Popek and Goldberg(1974)]{popek_formal_1974}
G.~J. Popek and R.~P. Goldberg.
\newblock Formal {Requirements} for {Virtualizable} {Third} {Generation}
  {Architectures}.
\newblock \emph{Communications of the ACM}, 17\penalty0 (7):\penalty0 412--421,
  July 1974.

\bibitem[Postel(1981)]{postel_internet_1981}
J.~Postel.
\newblock Internet {Protocol}.
\newblock Request for {Comments} 791, Internet Engineering Task Force (IETF),
  Defense Advanced Research Projects Agency (DARPA), Marina del Rey,
  California, Sept. 1981.

\bibitem[Price and Tucker(2004)]{price_solaris_2004}
D.~Price and A.~Tucker.
\newblock Solaris {Zones}: {Operating} {System} {Support} for {Consolidating}
  {Commercial} {Workloads}.
\newblock In \emph{Proceedings of the 18th {USENIX} {Conference} on {System}
  {Administration}}, {LISA} '04, pages 241--254, Berkeley, CA, USA, 2004.
  USENIX Association.

\bibitem[Priedhorsky and Randles(2017)]{priedhorsky_charliecloud:_2017}
R.~Priedhorsky and T.~Randles.
\newblock Charliecloud: {Unprivileged} {Containers} for {User}-defined
  {Software} {Stacks} in {HPC}.
\newblock In \emph{Proceedings of the {International} {Conference} for {High}
  {Performance} {Computing}, {Networking}, {Storage} and {Analysis}}, {SC} '17,
  pages 36:1--36:10, New York, NY, USA, 2017. ACM.

\bibitem[Raho et~al.(2015)Raho, Spyridakis, Paolino, and Raho]{raho_kvm_2015}
M.~Raho, A.~Spyridakis, M.~Paolino, and D.~Raho.
\newblock {KVM}, {Xen} and {Docker}: {A} performance analysis for {ARM} based
  {NFV} and cloud computing.
\newblock In \emph{2015 {IEEE} 3rd {Workshop} on {Advances} in {Information},
  {Electronic} and {Electrical} {Engineering} ({AIEEE})}, pages 1--8, Nov.
  2015.

\bibitem[Reshetova et~al.(2014)Reshetova, Karhunen, Nyman, and
  Asokan]{reshetova_security_2014}
E.~Reshetova, J.~Karhunen, T.~Nyman, and N.~Asokan.
\newblock Security of {OS}-{Level} {Virtualization} {Technologies}.
\newblock In K.~Bernsmed and S.~Fischer-H{\"u}bner, editors, \emph{Secure {IT}
  {Systems}}, Lecture {Notes} in {Computer} {Science}, pages 77--93. Springer
  International Publishing, 2014.

\bibitem[Ritchie(1980)]{ritchie_evolution_1980}
D.~Ritchie.
\newblock The {Evolution} of the {Unix} {Time}-{Sharing} {System}.
\newblock In \emph{Proceedings of a {Symposium} on {Language} {Design} and
  {Programming} {Methodology}}, volume~79 of \emph{Lecture {Notes} in
  {Computer} {Science}}, pages 25--36, London, UK, UK, 1980. Springer-Verlag.

\bibitem[Rizzo and Ranganathan(2018)]{rizzo_titan:_2018}
D.~Rizzo and P.~Ranganathan.
\newblock Titan: {Google}{\textquoteright}s {Root}-of-{Trust} {Security}
  {Silicon}.
\newblock In \emph{Proceedings of the {IEEE} {Hot} {Chips} {Symposium}},
  Cupertino, CA, Aug. 2018. IEEE.

\bibitem[Robin and Irvine(2000)]{robin_analysis_2000}
J.~S. Robin and C.~E. Irvine.
\newblock Analysis of the {Intel} {Pentium}{\textquoteright}s {Ability} to
  {Support} a {Secure} {Virtual} {Machine} {Monitor}.
\newblock In \emph{Proceedings of the 9th {USENIX} {Security} {Symposium}},
  pages 129--144, Denver, CO, 2000. USENIX Association.

\bibitem[Rochester(1955)]{rochester_computer_1955}
N.~Rochester.
\newblock The {Computer} and {Its} {Peripheral} {Equipment}.
\newblock In \emph{Papers and {Discussions} {Presented} at the the {November}
  7-9, 1955, {Eastern} {Joint} {AIEE}-{IRE} {Computer} {Conference}:
  {Computers} in {Business} and {Industrial} {Systems}}, {AIEE}-{IRE} '55
  ({Eastern}), pages 64--69, New York, NY, USA, 1955. ACM.

\bibitem[Rosenblum and Garfinkel(2005)]{rosenblum_virtual_2005}
M.~Rosenblum and T.~Garfinkel.
\newblock Virtual {Machine} {Monitors}: {Current} {Technology} and {Future}
  {Trends}.
\newblock \emph{Computer}, 38\penalty0 (5):\penalty0 39--47, May 2005.

\bibitem[Rushby(1981)]{rushby_design_1981}
J.~M. Rushby.
\newblock Design and {Verification} of {Secure} {Systems}.
\newblock In \emph{Proceedings of the {Eighth} {ACM} {Symposium} on {Operating}
  {Systems} {Principles}}, {SOSP} '81, pages 12--21, New York, NY, USA, 1981.
  ACM.

\bibitem[Saltzer and Schroeder(1975)]{saltzer_protection_1975}
J.~H. Saltzer and M.~D. Schroeder.
\newblock The protection of information in computer systems.
\newblock \emph{Proceedings of the IEEE}, 63\penalty0 (9):\penalty0 1278--1308,
  Sept. 1975.

\bibitem[Sapuntzakis et~al.(2002)Sapuntzakis, Chandra, Pfaff, Chow, Lam, and
  Rosenblum]{sapuntzakis_optimizing_2002}
C.~P. Sapuntzakis, R.~Chandra, B.~Pfaff, J.~Chow, M.~S. Lam, and M.~Rosenblum.
\newblock Optimizing the {Migration} of {Virtual} {Computers}.
\newblock \emph{SIGOPS Oper. Syst. Rev.}, 36\penalty0 (SI):\penalty0 377--390,
  Dec. 2002.

\bibitem[Schimunek et~al.(1999)Schimunek, Dupuche, Fung, Kirkdale, Myhra, and
  Stein]{schimunek_slicing_1999}
G.~Schimunek, D.~Dupuche, T.~Fung, P.~Kirkdale, E.~Myhra, and H.~Stein.
\newblock \emph{Slicing the {AS}/400 with {Logical} {Partitioning}: {A} {How}
  to {Guide}}.
\newblock IBM Corporation, Aug. 1999.

\bibitem[Schwarz et~al.(2018)Schwarz, Schwarzl, Lipp, and
  Gruss]{schwarz_netspectre:_2018}
M.~Schwarz, M.~Schwarzl, M.~Lipp, and D.~Gruss.
\newblock {NetSpectre}: {Read} {Arbitrary} {Memory} over {Network}.
\newblock \emph{arXiv:1807.10535 [cs]}, July 2018.

\bibitem[Seshadri et~al.(2007)Seshadri, Luk, Qu, and
  Perrig]{seshadri_secvisor:_2007}
A.~Seshadri, M.~Luk, N.~Qu, and A.~Perrig.
\newblock {SecVisor}: {A} {Tiny} {Hypervisor} to {Provide} {Lifetime} {Kernel}
  {Code} {Integrity} for {Commodity} {OSes}.
\newblock In \emph{Proceedings of {Twenty}-first {ACM} {SIGOPS} {Symposium} on
  {Operating} {Systems} {Principles}}, {SOSP} '07, pages 335--350, New York,
  NY, USA, 2007. ACM.

\bibitem[Shen et~al.(2019)Shen, Sun, Sela, Bagdasaryan, Delimitrou,
  Van~Renesse, and Weatherspoon]{shen_x-containers:_2019}
Z.~Shen, Z.~Sun, G.-E. Sela, E.~Bagdasaryan, C.~Delimitrou, R.~Van~Renesse, and
  H.~Weatherspoon.
\newblock X-{Containers}: {Breaking} {Down} {Barriers} to {Improve}
  {Performance} and {Isolation} of {Cloud}-{Native} {Containers}.
\newblock In \emph{Proceedings of the 24th {ACM} {International} {Conference}
  on {Architectural} {Support} for {Programming} {Languages} and {Operating}
  {Systems} ({ASPLOS} '19) [{Preprint}]}, page~15, Providence, RI, USA, Apr.
  2019. ACM.

\bibitem[Singh et~al.(2014)Singh, Bacon, Crowcroft, Madhavapeddy, Pasquier,
  Hon, and Millard]{singh_regional_2014}
J.~Singh, J.~Bacon, J.~Crowcroft, A.~Madhavapeddy, T.~Pasquier, W.~K. Hon, and
  C.~Millard.
\newblock Regional clouds: technical considerations.
\newblock Technical Report UCAM-CL-TR-863, University of Cambridge, Computer
  Laboratory, Nov. 2014.

\bibitem[Smith and Nair(2005)]{smith_architecture_2005}
J.~E. Smith and R.~Nair.
\newblock The architecture of virtual machines.
\newblock \emph{Computer}, 38\penalty0 (5):\penalty0 32--38, May 2005.

\bibitem[Soltesz et~al.(2007)Soltesz, P{\"o}tzl, Fiuczynski, Bavier, and
  Peterson]{soltesz_container-based_2007}
S.~Soltesz, H.~P{\"o}tzl, M.~E. Fiuczynski, A.~Bavier, and L.~Peterson.
\newblock Container-based {Operating} {System} {Virtualization}: {A}
  {Scalable}, {High}-performance {Alternative} to {Hypervisors}.
\newblock In \emph{Proceedings of the 2nd {ACM} {SIGOPS}/{EuroSys} {European}
  {Conference} on {Computer} {Systems} 2007}, pages 275--287, New York, NY,
  USA, 2007. ACM.

\bibitem[Soltis(2001)]{soltis_fortress_2001}
F.~G. Soltis.
\newblock \emph{Fortress {Rochester}: {The} {Inside} {Story} of the {IBM}
  {ISeries}}.
\newblock System iNetwork, 2001.

\bibitem[Souppaya et~al.(2017)Souppaya, Morello, and
  Scarfone]{souppaya_application_2017}
M.~Souppaya, J.~Morello, and K.~Scarfone.
\newblock Application container security guide.
\newblock Technical Report NIST SP 800-190, National Institute of Standards and
  Technology, Gaithersburg, MD, Sept. 2017.

\bibitem[Srodawa and Bates(1973)]{srodawa_efficient_1973}
R.~J. Srodawa and L.~A. Bates.
\newblock An efficient virtual machine implementation.
\newblock pages 43--73. ACM, Mar. 1973.

\bibitem[Sturgis(1973)]{sturgis_postmortem_1973}
H.~E. Sturgis.
\newblock \emph{A postmortem for a time sharing system}.
\newblock {PhD} {Thesis}, University of California at Berkeley, Berkeley, CA,
  June 1973.

\bibitem[Suda(2019)]{suda_allow_2019}
A.~Suda.
\newblock Allow running dockerd as a non-root user ({Rootless} mode), Feb.
  2019.
\newblock URL \url{https://github.com/moby/moby/pull/38050}.

\bibitem[Suda and Scrivano(2019)]{suda_rootless_2019}
A.~Suda and G.~Scrivano.
\newblock Rootless {Kubernetes}, Feb. 2019.
\newblock URL
  \url{https://fosdem.org/2019/schedule/event/containers_k8s_rootless/}.

\bibitem[Syed and Fernandez(2017)]{syed_container_2017}
M.~H. Syed and E.~B. Fernandez.
\newblock The {Container} {Manager} {Pattern}.
\newblock In \emph{Proceedings of the 22nd {European} {Conference} on {Pattern}
  {Languages} of {Programs}}, {EuroPLoP} '17, pages 28:1--28:9, New York, NY,
  USA, 2017. ACM.

\bibitem[Syed and Fernandez(2018)]{syed_reference_2018}
M.~H. Syed and E.~B. Fernandez.
\newblock A {Reference} {Architecture} for the {Container} {Ecosystem}.
\newblock In \emph{Proceedings of the 13th {International} {Conference} on
  {Availability}, {Reliability} and {Security}}, {ARES} 2018, pages 31:1--31:6,
  New York, NY, USA, 2018. ACM.

\bibitem[Szefer et~al.(2011)Szefer, Keller, Lee, and
  Rexford]{szefer_eliminating_2011}
J.~Szefer, E.~Keller, R.~B. Lee, and J.~Rexford.
\newblock Eliminating the {Hypervisor} {Attack} {Surface} for a {More} {Secure}
  {Cloud}.
\newblock In \emph{Proceedings of the 18th {ACM} {Conference} on {Computer} and
  {Communications} {Security}}, {CCS} '11, pages 401--412, New York, NY, USA,
  2011. ACM.

\bibitem[Szekeres et~al.(2013)Szekeres, Payer, {Tao Wei}, and
  Song]{szekeres_sok:_2013}
L.~Szekeres, M.~Payer, {Tao Wei}, and D.~Song.
\newblock {SoK}: {Eternal} {War} in {Memory}.
\newblock In \emph{2013 {IEEE} {Symposium} on {Security} and {Privacy}}, pages
  48--62, Berkeley, CA, May 2013. IEEE.

\bibitem[Van~Bulck et~al.(2018)Van~Bulck, Minkin, Weisse, Genkin, Kasikci,
  Piessens, Silberstein, Wenisch, Yarom, and
  Strackx]{van_bulck_foreshadow:_2018}
J.~Van~Bulck, M.~Minkin, O.~Weisse, D.~Genkin, B.~Kasikci, F.~Piessens,
  M.~Silberstein, T.~F. Wenisch, Y.~Yarom, and R.~Strackx.
\newblock Foreshadow: {Extracting} the {Keys} to the {Intel} {SGX} {Kingdom}
  with {Transient} {Out}-of-{Order} {Execution}.
\newblock In \emph{27th {USENIX} {Security} {Symposium} ({USENIX} {Security}
  18)}, pages 991--1008, Baltimore, MD, Aug. 2018. USENIX Association.

\bibitem[Vasudevan et~al.(2010)Vasudevan, McCune, Qu, Van~Doorn, and
  Perrig]{vasudevan_requirements_2010}
A.~Vasudevan, J.~M. McCune, N.~Qu, L.~Van~Doorn, and A.~Perrig.
\newblock Requirements for an {Integrity}-protected {Hypervisor} on the x86
  {Hardware} {Virtualized} {Architecture}.
\newblock In \emph{Proceedings of the 3rd {International} {Conference} on
  {Trust} and {Trustworthy} {Computing}}, {TRUST}'10, pages 141--165, Berlin,
  Heidelberg, 2010. Springer-Verlag.

\bibitem[Verma et~al.(2015)Verma, Pedrosa, Korupolu, Oppenheimer, Tune, and
  Wilkes]{verma_large-scale_2015}
A.~Verma, L.~Pedrosa, M.~Korupolu, D.~Oppenheimer, E.~Tune, and J.~Wilkes.
\newblock Large-scale {Cluster} {Management} at {Google} with {Borg}.
\newblock In \emph{Proceedings of the {Tenth} {European} {Conference} on
  {Computer} {Systems}}, {EuroSys} '15, pages 18:1--18:17, New York, NY, USA,
  2015. ACM.

\bibitem[Waldspurger(2002)]{waldspurger_memory_2002}
C.~A. Waldspurger.
\newblock Memory {Resource} {Management} in {VMware} {ESX} {Server}.
\newblock \emph{SIGOPS Oper. Syst. Rev.}, 36\penalty0 (SI):\penalty0 181--194,
  Dec. 2002.

\bibitem[Wang et~al.(2012)Wang, Wu, Grace, and Jiang]{wang_isolating_2012}
Z.~Wang, C.~Wu, M.~Grace, and X.~Jiang.
\newblock Isolating {Commodity} {Hosted} {Hypervisors} with {HyperLock}.
\newblock In \emph{Proceedings of the 7th {ACM} {European} {Conference} on
  {Computer} {Systems}}, {EuroSys} '12, pages 127--140, New York, NY, USA,
  2012. ACM.

\bibitem[Watson et~al.(2010)Watson, Anderson, Laurie, and
  Kennaway]{watson_capsicum:_2010}
R.~Watson, J.~Anderson, B.~Laurie, and K.~Kennaway.
\newblock Capsicum: {Practical} {Capabilities} for {UNIX}.
\newblock In \emph{Proceedings of the 19th {USENIX} {Security} {Symposium}},
  volume~19, Washington, DC, USA, Aug. 2010. ACM Press.

\bibitem[Watson et~al.(2012{\natexlab{a}})Watson, Neumann, Woodruff, Anderson,
  Anderson, Dave, Laurie, Moore, Murdoch, Paeps, Roe, and
  Saidi]{watson_cheri:_2012}
R.~Watson, P.~Neumann, J.~Woodruff, J.~Anderson, R.~Anderson, N.~Dave,
  B.~Laurie, S.~W. Moore, S.~J. Murdoch, P.~Paeps, M.~Roe, and H.~Saidi.
\newblock {CHERI}: a research platform deconflating hardware virtualization and
  protection.
\newblock In \emph{Unpublished workshop paper for
  {RESoLVE}{\textquoteright}12}, London, UK, Mar. 2012{\natexlab{a}}.

\bibitem[Watson et~al.(2012{\natexlab{b}})Watson, Anderson, Laurie, and
  Kennaway]{watson_taste_2012}
R.~N.~M. Watson, J.~Anderson, B.~Laurie, and K.~Kennaway.
\newblock A {Taste} of {Capsicum}: {Practical} {Capabilities} for {UNIX}.
\newblock \emph{Communications of the ACM}, 55\penalty0 (3):\penalty0 97--104,
  Mar. 2012{\natexlab{b}}.

\bibitem[Watson et~al.(2014)Watson, Neumann, Woodruff, Anderson, Chisnall,
  Davis, Laurie, Moore, Murdoch, and Roe]{watson_capability_2014}
R.~N.~M. Watson, P.~G. Neumann, J.~Woodruff, J.~Anderson, D.~Chisnall,
  B.~Davis, B.~Laurie, S.~W. Moore, S.~J. Murdoch, and M.~Roe.
\newblock Capability {Hardware} {Enhanced} {RISC} {Instructions}: {CHERI}
  {Instruction}-set architecture.
\newblock Technical Report UCAM-CL-TR-864, University of Cambridge, Computer
  Laboratory, Dec. 2014.

\bibitem[Watson et~al.(2015)Watson, Woodruff, Neumann, Moore, Anderson,
  Chisnall, Dave, Davis, Gudka, Laurie, Murdoch, Norton, Roe, Son, and
  Vadera]{watson_cheri:_2015}
R.~N.~M. Watson, J.~Woodruff, P.~G. Neumann, S.~W. Moore, J.~Anderson,
  D.~Chisnall, N.~Dave, B.~Davis, K.~Gudka, B.~Laurie, S.~J. Murdoch,
  R.~Norton, M.~Roe, S.~Son, and M.~Vadera.
\newblock {CHERI}: {A} {Hybrid} {Capability}-{System} {Architecture} for
  {Scalable} {Software} {Compartmentalization}.
\newblock In \emph{2015 {IEEE} {Symposium} on {Security} and {Privacy}}, pages
  20--37, May 2015.

\bibitem[Watson et~al.(2018)Watson, Woodruff, Roe, Moore, and
  Neumann]{watson_capability_2018}
R.~N.~M. Watson, J.~Woodruff, M.~Roe, S.~W. Moore, and P.~G. Neumann.
\newblock Capability {Hardware} {Enhanced} {RISC} {Instructions} ({CHERI}):
  {Notes} on the {Meltdown} and {Spectre} {Attacks}.
\newblock Technical Report UCAM-CL-TR-916, University of Cambridge, Computer
  Laboratory, Feb. 2018.

\bibitem[Weisse et~al.(2018)Weisse, Bulck, Minkin, Genkin, Kasikci, Piessens,
  Silberstein, Strackx, Wenisch, and Yarom]{weisse_foreshadow-ng:_2018}
O.~Weisse, J.~V. Bulck, M.~Minkin, D.~Genkin, B.~Kasikci, F.~Piessens,
  M.~Silberstein, R.~Strackx, T.~F. Wenisch, and Y.~Yarom.
\newblock Foreshadow-{NG}: {Breaking} the {Virtual} {Memory} {Abstraction} with
  {Transient} {Out}-of-{Order} {Execution}.
\newblock Technical report, Aug. 2018.

\bibitem[Whitaker et~al.(2002{\natexlab{a}})Whitaker, Shaw, and
  Gribble]{whitaker_denali:_2002-1}
A.~Whitaker, M.~Shaw, and S.~Gribble.
\newblock Denali: {Lightweight} {Virtual} {Machines} for {Distributed} and
  {Networked} {Applications}.
\newblock Technical report, University of Washington, 2002{\natexlab{a}}.

\bibitem[Whitaker et~al.(2002{\natexlab{b}})Whitaker, Shaw, and
  Gribble]{whitaker_denali:_2002}
A.~Whitaker, M.~Shaw, and S.~D. Gribble.
\newblock Denali: {A} {Scalable} {Isolation} {Kernel}.
\newblock In \emph{Proceedings of the 10th {Workshop} on {ACM} {SIGOPS}
  {European} {Workshop}}, pages 10--15, New York, NY, USA, 2002{\natexlab{b}}.
  ACM.

\bibitem[Whitaker et~al.(2002{\natexlab{c}})Whitaker, Shaw, and
  Gribble]{whitaker_scale_2002}
A.~Whitaker, M.~Shaw, and S.~D. Gribble.
\newblock Scale and {Performance} in the {Denali} {Isolation} {Kernel}.
\newblock \emph{SIGOPS Oper. Syst. Rev.}, 36\penalty0 (SI):\penalty0 195--209,
  Dec. 2002{\natexlab{c}}.

\bibitem[Wilkes(1968)]{wilkes_time-sharing_1968}
M.~V. Wilkes.
\newblock \emph{Time-{Sharing} {Computer} {Systems}}.
\newblock Number~5 in {MacDonald} {Computer} {Monographs}. MacDonald \& Co., 2
  edition, 1968.

\bibitem[Wilkes(1979)]{wilkes_cambridge_1979}
M.~V. Wilkes.
\newblock \emph{The {Cambridge} {CAP} {Computer} and {Its} {Operating} {System}
  ({Operating} and {Programming} {Systems} {Series})}.
\newblock North-Holland Publishing Co., Amsterdam, The Netherlands, 1979.

\bibitem[Wilkes and Willis(1956)]{wilkes_magnetic-tape_1956}
M.~V. Wilkes and D.~W. Willis.
\newblock A magnetic-tape auxiliary storage system for the {EDSAC}.
\newblock \emph{Proceedings of the IEE - Part B: Radio and Electronic
  Engineering}, 103\penalty0 (2):\penalty0 337--345, Apr. 1956.

\bibitem[Williams and Koller(2016)]{williams_unikernel_2016}
D.~Williams and R.~Koller.
\newblock Unikernel {Monitors}: {Extending} {Minimalism} {Outside} of the
  {Box}.
\newblock In \emph{8th {USENIX} {Workshop} on {Hot} {Topics} in {Cloud}
  {Computing} ({HotCloud} 16)}, page~6, Denver, CO, 2016. USENIX Association.

\bibitem[Williams et~al.(2018{\natexlab{a}})Williams, Koller, Lucina, and
  Prakash]{williams_unikernels_2018}
D.~Williams, R.~Koller, M.~Lucina, and N.~Prakash.
\newblock Unikernels {As} {Processes}.
\newblock In \emph{Proceedings of the {ACM} {Symposium} on {Cloud}
  {Computing}}, {SoCC} '18, pages 199--211, New York, NY, USA,
  2018{\natexlab{a}}. ACM.

\bibitem[Williams et~al.(2018{\natexlab{b}})Williams, Koller, and
  Lum]{williams_say_2018}
D.~Williams, R.~Koller, and B.~Lum.
\newblock Say {Goodbye} to {Virtualization} for a {Safer} {Cloud}.
\newblock In \emph{10th {USENIX} {Workshop} on {Hot} {Topics} in {Cloud}
  {Computing} ({HotCloud} 18)}, Boston, MA, 2018{\natexlab{b}}. USENIX
  Association.

\bibitem[Woodruff et~al.(2014)Woodruff, Watson, Chisnall, Moore, Anderson,
  Davis, Laurie, Neumann, Norton, and Roe]{woodruff_cheri_2014}
J.~Woodruff, R.~N. Watson, D.~Chisnall, S.~W. Moore, J.~Anderson, B.~Davis,
  B.~Laurie, P.~G. Neumann, R.~Norton, and M.~Roe.
\newblock The {CHERI} {Capability} {Model}: {Revisiting} {RISC} in an {Age} of
  {Risk}.
\newblock In \emph{Proceeding of the 41st {Annual} {International} {Symposium}
  on {Computer} {Architecuture}}, {ISCA} '14, pages 457--468, Piscataway, NJ,
  USA, 2014. IEEE Press.

\bibitem[Wright et~al.(2002)Wright, Cowan, Smalley, Morris, and
  Kroah-Hartman]{wright_linux_2002}
C.~Wright, C.~Cowan, S.~Smalley, J.~Morris, and G.~Kroah-Hartman.
\newblock Linux {Security} {Module} {Framework}.
\newblock In \emph{Proceedings of the {Ottawa} {Linux} {Symposium}}, pages
  604--617, Ottowa, Canada, June 2002.

\bibitem[Wulf et~al.(1974)Wulf, Cohen, Corwin, Jones, Levin, Pierson, and
  Pollack]{wulf_hydra:_1974}
W.~Wulf, E.~Cohen, W.~Corwin, A.~Jones, R.~Levin, C.~Pierson, and F.~Pollack.
\newblock {HYDRA}: {The} {Kernel} of a {Multiprocessor} {Operating} {System}.
\newblock \emph{Commun. ACM}, 17\penalty0 (6):\penalty0 337--345, June 1974.

\bibitem[Wulf et~al.(1981)Wulf, Levin, and Harbison]{wulf_hydra-c._1981}
W.~A. Wulf, R.~Levin, and S.~P. Harbison.
\newblock \emph{{HYDRA}-{C}. {Mmp}: {An} {Experimental} {Computer} {System}}.
\newblock McGraw-Hill, 1981.

\bibitem[Xavier et~al.(2013)Xavier, Neves, Rossi, Ferreto, Lange, and
  Rose]{xavier_performance_2013}
M.~G. Xavier, M.~V. Neves, F.~D. Rossi, T.~C. Ferreto, T.~Lange, and C.~A.
  F.~D. Rose.
\newblock Performance {Evaluation} of {Container}-{Based} {Virtualization} for
  {High} {Performance} {Computing} {Environments}.
\newblock In \emph{2013 21st {Euromicro} {International} {Conference} on
  {Parallel}, {Distributed}, and {Network}-{Based} {Processing}}, pages
  233--240, Feb. 2013.

\bibitem[Zhai et~al.(2016)Zhai, Yin, Chase, Ristenpart, and
  Swift]{zhai_cqstr:_2016}
Y.~Zhai, L.~Yin, J.~Chase, T.~Ristenpart, and M.~Swift.
\newblock {CQSTR}: {Securing} {Cross}-{Tenant} {Applications} with {Cloud}
  {Containers}.
\newblock In \emph{Proceedings of the {Seventh} {ACM} {Symposium} on {Cloud}
  {Computing}}, {SoCC} '16, pages 223--236, New York, NY, USA, 2016. ACM.

\end{thebibliography}

\end{document}